\documentclass{emulateapj}
\usepackage{graphicx,graphics,amsmath}
\usepackage{natbib}
\usepackage{color}
\usepackage[normalem]{ulem}
\usepackage[plainpages=false, colorlinks=true, anchorcolor=blue,
linkcolor=blue, citecolor=blue, bookmarks=false]{hyperref}
\begin{document}
\title{A theoretical study of the build-up of the Sun's polar magnetic field by using a 3D kinematic dynamo model}
\shorttitle{Build-up of Solar Polar Field} %

\shortauthors{Hazra, Choudhuri \& Miesch}
\author{Gopal Hazra\altaffilmark{1,2}, Arnab Rai Choudhuri\altaffilmark{1} and Mark S. Miesch\altaffilmark{3}}
\affil{$^1$Department of Physics, Indian Institute of Science, Bangalore, India 560012} %
\affil{$^2$Indian Institute of Astrophysics, Bangalore, India 560034} %
\affil{$^3$ High Altitude Observatory, National Center for Atmospheric Research, Boulder, CO 80301, USA} %
\email{ghazra@physics.iisc.ernet.in,}
\email{arnab@physics.iisc.ernet.in \& miesch@ucar.edu}

\begin{abstract}
We develop a three-dimensional kinematic self-sustaining model of the solar
dynamo in which the poloidal field generation is from tilted bipolar sunspot
pairs placed on the solar surface above regions of strong toroidal field by using
the SpotMaker algorithm, and then the transport of this poloidal field to the
tachocline is primarily caused by turbulent diffusion. We obtain a dipolar
solution within a certain range of parameters. We use this model to study the
build-up of the polar magnetic field and show that some insights obtained from
surface flux transport (SFT) models have to be revised.  We present results
obtained by putting a single bipolar sunspot pair in a hemisphere and two symmetrical
sunspot pairs in two hemispheres.We find that the polar fields produced by them
disappear due to the upward advection of poloidal flux at low latitudes, which emerges
as oppositely-signed radial flux and which is then advected poleward by the meridional flow.
We also study the effect that a large sunspot pair, violating Hale's polarity law would
have on the polar field. We find that there would be some effect---especially if
the anti-Hale pair appears at high latitudes in the mid-phase of the cycle---though
the effect is not very dramatic.
\end{abstract}
\section{Introduction}

The flux transport dynamo model, which started being developed from the 1990s
\citep{WSN91,CSD95,Durney95}, has emerged as
an attractive theoretical model for explaining the solar cycle and 
has been extensively reviewed by several authors \citep{chou11,Charbonneau14,Karakreview14}.
In any dynamo model, the toroidal magnetic field
is generated from the poloidal field by the differential rotation, which has
now been mapped by helioseismology \citep{Thompson96}. The distinctive features of
the B-L flux transport dynamo model are that the meridional circulation plays a crucial
role in this model and the poloidal magnetic field is generated by the Babcock--Leighton
(BL) process involving the decay of tilted bipolar sunspots.  Bipolar sunspots
are assumed to form due to the buoyant rise of the toroidal magnetic flux
through the convection zone \citep{Parker55b} and their tilts result from the action of the
Coriolis force on the rising flux tubes \citep{chou89,Dsilva93, Fan93}
leading to Joy's law \citep{Hale19}. When a tilted pair of bipolar sunspots
decays, turbulent diffusion spreads the magnetic flux to produce a poloidal magnetic
component \citep{Bab61,Leighton64}. An over-all poloidal field develops
from the contributions due to many bipolar sunspots and 
is advected to the poles by the meridional circulation,
which is poleward in the upper layers of the convection zone.  The polar magnetic
field of the Sun is built up in this process.

The BL process---which involves the production of tilted bipolar sunspot pairs
and the generation of the poloidal field from their decay---is an inherently 3D
process and can be modeled in 2D only through drastically simplified crude approximations
\citep{CH16}. Still an understanding of how the poloidal field builds
up by the BL process historically came from two distinct classes of 2D theoretical
models: the 2D flux transport dynamo model and the surface flux transport (SFT)
model.  In the 2D flux transport dynamo model, we average over the azimuthal
direction $\phi$ and solve the axisymmetric dynamo equation in the $r$-$\theta$
plane. On the other hand, in the SFT model, we focus our attention
only on the $B_r$ component of the magnetic field
at the solar surface spanned by the $\theta$-$\phi$
coordinates and study its evolution on this surface under the joint action  of diffusion, meridional
circulation and differential rotation. Neither of these approaches provides a fully
satisfactory depiction of the BL process and each approach has its own limitations.

If a tilted bipolar sunspot pair at the solar surface is averaged over the 
azimuthal direction $\phi$, then we get two rings of opposite magnetic polarity
at slightly different latitudes. \citet{Durney95,Durney97} advocated the development
of the flux transport dynamo model by using such double rings as the source of the
poloidal component. However, a more popular approach has been to introduce an $\alpha$-coefficient
reminiscent of the $\alpha$-effect of the mean field dynamo theory (\citet{Parker55a,SKR66,
Chou98}, Chapter 16), although this now has a completely different interpretation.
The source term of the poloidal field is taken as the product of this $\alpha$-coefficient
and the toroidal field that has risen from the tachocline due to magnetic buoyancy.
\citet{CH16} review how different authors achieve this, with references
to the original papers. \citet{Nandy01} showed that the double ring approach
and the treatment through $\alpha$-coefficient give qualitatively similar results,
although \citet{Munoz10} argued that the double  ring approach is more
realistic. In any case, the 2D kinematic dynamo models do not give a detailed picture
of how the poloidal field builds up from the contributions of many individual bipolar
sunspot pairs, since such pairs get smeared over when we average over the azimuthal
direction. Also, as most of these dynamo models rely on a mean field approach,
flux tubes or sunspots are not handled properly in these models \citep{Chou03}.

Starting from the pioneering work of \citet{Wang89a,Wang89b}, the surface flux
transport (SFT) model has been made more sophisticated in several recent studies
\citep{van98,Schrijver02,Baumann04,Baumann06,Cameron10, Jiang14, UH1_14, Jiang15}. 
In this model, recently reviewed by \citet{Jiang_review15}, one can study in detail
how individual sunspot pairs contribute in building up the poloidal field and can address
such questions as to how this process depends on such factors as the latitudinal positions
of the sunspot pairs and the distribution in their tilt angles.  The main limitation
of this model is that several important aspects of physics get left out by ignoring the
vectorial nature of the magnetic field and by not including any subsurface
processes.  By studying the time evolution of an axisymmetric poloidal field,
\citet{DC94,DC95} and \citet{CD99} showed that the subduction of the poloidal field by the meridional circulation sinking
underneath the surface at the polar region plays an important role in the dynamics
of the magnetic field.  Since this process cannot be included in the SFT
models, flux of $B_r$ tends to get piled up in the polar regions
and has to be neutralized by flux of the opposite sign advected there. If additional flux
of the opposite sign is not brought there, then the polar field may reach an asymptotic
value as seen in Figure~6 of \citet{Jiang14}. When one tries to model several
successive cycles through an SFT model, one may get a `secular
drift' of the polar field if the flux of the succeeding cycle is unable to properly
neutralize the polar flux of the preceding cycle, as seen in Figure~1 of \citet{Baumann06}.
A way of fixing this problem proposed by \citet{Baumann06}
involves adding an ad hoc decay term corresponding to the radial diffusion not
included in the SFT model. 
In spite of the tremendously important historical role the SFT model has
played in elucidating the BL process, this model has the inherent limitation that
it cannot adequately handle the magnetic field dynamics in the Sun's polar region. 

We believe that the next step forward is the 3D kinematic flux transport dynamo model.  
In this model, the fluid motions (differential rotation, meridional circulation) are specified
and the evolution of the magnetic field is calculated in 3D. Such a model has the
promise of incorporating the attractive features of both the 2D flux transport dynamo
model and the SFT model, while being free from the limitations of both
these models. It can handle the BL process much more realistically than the 2D flux
transport dynamo model, where we average over the azimuthal direction and cannot include
tilted bipolar sunspots properly.  On the other hand, this model incorporates the vectorial
nature of the magnetic field and the subsurface processes which are left out in the
SFT models.

Efforts of constructing 3D kinematic flux transport dynamo models began only
within the last few years. In a landmark paper,
\citet{YM13} developed a method of treating the buoyant rise
of a flux tube in their 3D dynamo model by simultaneously applying a radially outward
and a vortical velocity to a localized part of an azimuthal flux tube at the bottom of 
the convection zone. Although they did not present a self-excited dynamo solution, 
they simulated a solar cycle by incorporating bipolar sunspot eruptions by this method at the
actual locations where bipolar sunspots were observationally seen.
\citet{MD14} succeeded in producing a self-excited dynamo by identifying the locations (in latitude
and longitude) at the bottom of the convection zone where the toroidal field was
the strongest (in the theoretical model)
and then putting tilted bipolar sunspot pairs above those locations 
with loop-like magnetic structures both above and below the surface (by using an
algorithm which they named SpotMaker). More details of this model have been
given by \citet{MT16}. After a part of the toroidal flux tube
rises to the surface to produce bipolar sunspots, the magnetic field underneath these
sunspots has to be detached at some stage from the bottom of the convection zone before
the magnetic flux of the sunspots is dispersed freely by diffusion and carried poleward
by the meridional circulation \citep{Longcope02}. Our lack of understanding
of this process is the main difficulty in constructing realistic 3D dynamo models
at the present time.  Presumably, the approach of \citet{YM13}
captures the physics of the early phase soon after the bipolar sunspots emerge, whereas
the approach of \citet{MD14} is more appropriate for the later phase
when the magnetic field below the sunspot pairs has become detached from the bottom
of the convection zone.

The aim of the present paper is to use a modified version
of the model of \citet{MD14} to
study the build-up of the Sun's polar magnetic field by the BL process in more detail.
The model of \citet{MD14} uses values of parameters (such as turbulent
diffusion) which are probably not very realistic for the Sun.  We use the
same dynamo code named as STABLE (i.e.\ Surface flux Transport And Babcock LEighton Model)
to first construct a model of the solar dynamo based
on more realistic values of parameters and then use this model for our study.
Since the BL process has been studied most extensively by the SFT
models, we especially address the question whether the insights gained about various
aspects of this process from the SFT models are borne out by
the 3D model or have to be revised significantly. We shall see that the
accumulation of magnetic flux at the poles seen in the SFT models does
not occur when the low-latitude advection and emergence of oppositely-signed
radial flux is taken into account. Thus, the problem of `secular drift' is 
automatically eliminated. One insight from the SFT models is that the fluxes of
leading sunspots at lower latitudes get canceled across the equator and the fluxes 
from the following sunspots are then advected to the poles, building up a dipole moment of the Sun.
We shall see that this insight also will have to be modified significantly.  SFT models indicate that
even a few large sunspot pairs with anti-Hale or wrong polarity (i.e.\ opposite of what is expected
of sunspot pairs in that cycle) may have significant effect on the polar field \citep{Jiang15}.
We shall be able to study this effect more realistically
in our 3D model.

After discussing the mathematical formulation of the problem in the next section,
the standard model of the solar dynamo which we shall use is presented in \S~3. Then
the build-up of the polar field is studied in \S~4, while the effects of large anti-Hale
sunspot pairs are discussed in \S~5.  Our conclusions are summarized in \S~6.

\section{Mathematical Formulation}
In this section, we explain the basic formulation of the STABLE 
dynamo model which is first reported in \citet{MD14} and in more detail in \citet{MT16}. 
This model is a 3D generalization of the pre-existing axisymmetric 2D flux transport dynamo
models and it solves the induction equation in full 3D rotating spherical shell with radius
ranges from $r = 0.69R$ to $r = R$ of the Sun:

\begin{equation}
\label{induction1}
\frac{\partial {\bf B}}{\partial t} = \nabla \times ({\bf v}\times{\bf B} -\eta_t \nabla\times {\bf B})
+ S(\theta,\phi,t)
\end{equation}
where ${\bf v}$ is the velocity field, $\eta_t(r)$ is the turbulent diffusion in the solar 
convection zone, and $S(\theta,\phi,t)$ is the source function which captures the
effect of the BL mechanism. As we shall discuss in detail later, the source function
$S(\theta,\phi,t)$ is of the nature of an impulsive forcing term which becomes
non-zero only at the instants when we allow a bipolar sunspot pair to be put at
the solar surface. Though this model is fully 3D and no axisymmetric assumption is considered, 
still this model is kinematic and we provide the velocity field motivated from  
helioseismology and observations. We solve equation (\ref{induction1}) using Anelastic 
Spherical Harmonic (ASH) code \citep{Miesch_et_al_00,BMT04}. ASH is a well-established 
pseudospectral code which has been used extensively for 3D solar and stellar convection simulations, 
instabilities, tachocline confinement and many other aspects of solar and 
stellar internal dynamics. The ASH code has capability to solve the velocity equation and 
magnetic induction equation together but for our kinematic model we bypass the velocity 
equation solver and only solve the induction equation by providing observationally motivated velocity fields.  
The version of the code used for 3D kinematic dynamo modeling is named 
STABLE (i.e.\ Surface flux Transport And Babcock LEighton Model).

\begin{figure}[!htbp]
\centerline{\includegraphics[width=0.4\textwidth,clip=]{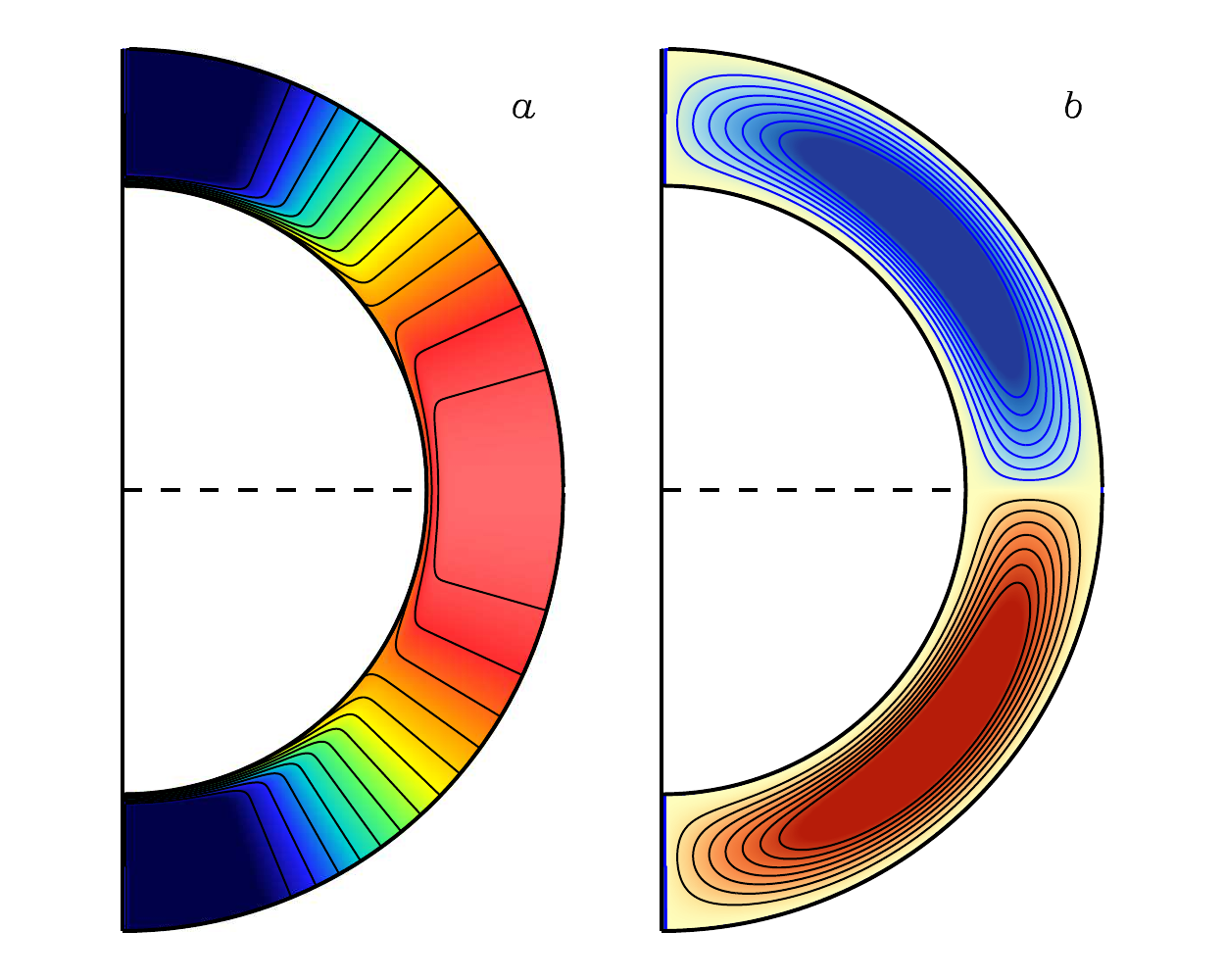}} 
\caption{(a). Differential rotation profile with color table ranging from 350-480 nHz from blue to red; (b) streamlines for the meridional flow. Blue and red contours show the poleward flow at the surface and an equatorward flow at the bottom of the convection zone in northern and southern hemisphere respectively. The amplitude of the meridional circulation is taken as 20.40 m/s on the surface and 1.64 m/s at the lower convection zone.}
\label{fig:MC}
\end{figure}

\begin{figure}[!htbp]
\centerline{\includegraphics[width=0.5\textwidth,clip=]{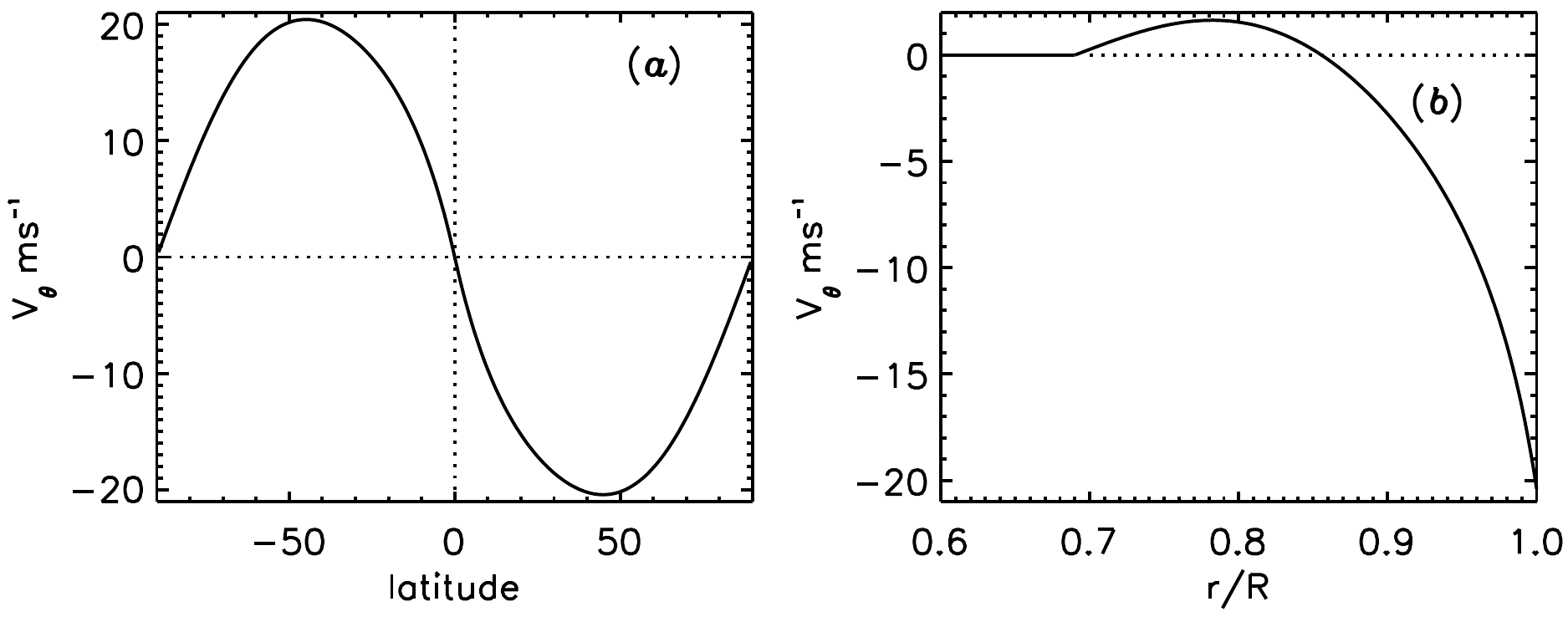}} 
\caption{Variation of latitudinal component of meridional circulation
$V_\theta$ with latitude on the surface (a) and with radius at $45^{\circ}$ latitude (b).}
\label{fig:vtheta}
\end{figure}

The mean velocity field ${\bf v}$ in the Sun can be written as the summation of the part from differential 
rotation $\Omega$ and the meridional circulation $v_p$. Whereas the exact differential 
rotation is mapped from helioseismology quite well \citep{Schou98}, the structure of the
meridional circulation in the solar convection zone is still under study. Recently, different 
helioseismology groups have reported substantially different structures of the meridional circulation in 
the solar convection zone \citep{Schad13, Zhao13, RA15}. In response to these observational claims, 
\citet{HKC14} carried on calculations with different types of 
meridional circulation structure and showed that flux transport dynamo works quite well as long as 
there is an equatorward flow at the bottom of the convection zone. Recently, \citet{KC16} 
showed that in the presence of appropriate profile of downward pumping in the solar convection zone, 
the flux transport solar dynamo works with even shallow meridional circulation. Since there
is still no compelling reason \citep{RA15} to give up the simple
single-cell profile of meridional circulation used by many previous authors (\citet{CNC04}, \citet{MD14}),
we use a single cell profile of the meridional circulation having 
a poleward flow at the surface and an equatorward return flow at the bottom of the convection 
zone. The stream function corresponding to the meridional circulation which we use here is
\begin{eqnarray}
\label{eq:psi}
\psi r \sin \theta = \psi_0 (r - R_p) \sin \left[ \frac{\pi (r - R_p)}{(R -R_p)} \right]\{ 1 - e^{- \beta_1 r\theta^{\epsilon}}\}\nonumber \\
\times\{1 - e^{\beta_2 r(\theta - \pi/2)} \} e^{-((r -r_0)/\Gamma)^2} ~~~~
\end{eqnarray}\\
with
$\beta_1 = 0.3 \times 10^{-10}~cm^{-1}, \beta_2 = 0.5 \times 10^{-10}~ cm^{-1}$, $\epsilon = 2.0000001$, $r_0 = (R - R_b)/4.0$, $\Gamma =
3.5 \times 10^{10}$ cm, $R_p = 0.69R$.
The value of $\psi_0$ determines the amplitude of the meridional circulation. On
taking $\psi_0 = 12.0$, the poleward flow near the surface at mid-latitudes peaks around $v_0=20.40$ m s$^{-1}$.
The contour plot for the meridional circulation is shown in 
figure~\ref{fig:MC}(b) and the variation of $V_\theta$ with latitude on the surface and 
variation $V_\theta$ with radius at mid-latitude $(45^{\circ})$ are shown in figure~\ref{fig:vtheta}(a) and \ref{fig:vtheta}(b) respectively.
For differential rotation we have used the analytical formula given
in \citep{DC99} which is a good fit to the observational data (Figure~\ref{fig:MC}(a)).

Turbulent diffusivity is another important parameter. After the BL process generates
the poloidal field near the solar surface, it has to reach the tachocline where the differential
rotation acts on it to produce the toroidal field.  This can happen in two ways. The poloidal
field may first be advected by the meridional circulation to the pole and then underneath
the surface to the mid-latitude tachocline from where the first sunspots of cycle rise.
The time scale for this is close to 20 years for a reasonable profile of the meridional circulation.
The second possibility is that the poloidal field diffuses from the surface to the bottom
of the convection zone to be acted upon by the differential rotation of the tachocline. 
The Green's function for the diffusion equation suggests that the diffusion
time across a length $L$ is $L^2/4 \eta_t$ (see, for example, \citet{Parker79}, p.\ 32). If
the turbulent diffusivity within the convection zone is assumed to be $5 \times 10^{11}$
cm$^2$ s$^{-1}$ as we shall do, then this diffusion time
turns out to be about 7 years if we take $L$ to be the thickness of the convection zone.
The value of the turbulent diffusivity determines whether the poloidal field
is transported across the convection zone primarily by meridional circulation or by
turbulent diffusion, and the behavior of the dynamo is very different in the two
situations \citep{Jiang07,Yeates08}. Over the years, we have got
more and more evidence that the turbulent diffusivity has to be sufficiently high to
make the poloidal field transport primarily by diffusion in order to explain many aspects
of the solar cycle, such as the dipolar parity \citep{CNC04,Hotta10}, the lack of significant hemispheric asymmetry
\citep{CC06,GoelChou09}, the observed correlation between the polar field at the cycle
minimum and the strength of the next cycle \citep{Jiang07}, the Waldmeier effect \citep{karakChou11}.
Such a value of turbulent diffusivity is also consistent with mixing length arguments
(\citet{Parker79}, p.\ 629) and the theory of mean flows \citep{miesc12b}.

\begin{figure}
\centerline{\includegraphics[width=0.45\textwidth,clip=]{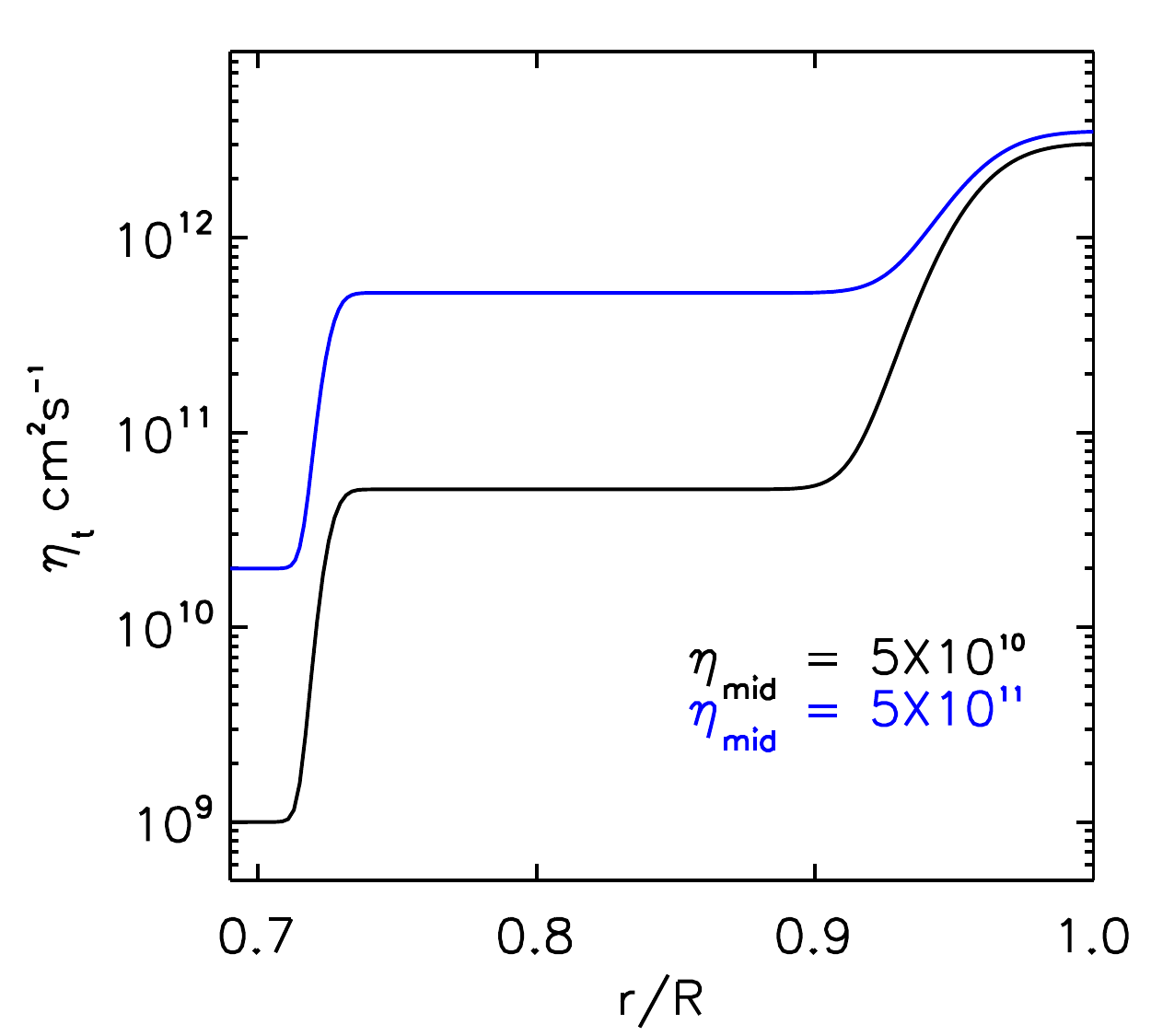}} 
\caption{High diffusivity profile used for most of our simulation is shown in blue solid line 
and the profile used for advection-dominated regime is shown in black solid line.}
\label{diff}
\end{figure}

In most of the SFT models, a constant diffusivity $(2.5-3.0) \times 10^{12}$ cm$^2$s$^{-1}$ on the 
surface is used \citep{Jiang_review15}. The turbulent diffusivity is expected to be less within the
convection zone and falls drastically at its bottom where convection is less vigorous.
Except when stated explicitly otherwise, the calculations of this paper assume the diffusivity
to be given by
\begin{eqnarray}
\eta_t = \eta_c + \frac{\eta_{mid}}{2}\left[1 + {\rm erf}\left(2\frac{r-r_{da}}{d_a}\right)\right] \nonumber \\
~~~+ \frac{\eta_{top}}{2}\left[1+{\rm erf}\left(2\frac{r-r_{db}}{d_b}\right)\right] 
\end{eqnarray} 
where $\eta_c = 2\times10^{10}$ cm$^2$ s$^{-1}$, $\eta_{mid}= 5\times10^{11}$ 
cm$^2s^{-1}$,$r_{da} = 0.725 R$, $r_{db} = 0.956R$, and $d_b = 0.05R$.
In Figure~\ref{diff} we have shown the diffusivity profile by the blue solid line. For 
comparison, the diffusivity profile used by \citet{MD14}
is shown by the black solid line.  It may be noted that some groups \citep{DC99,DG06},
over the years, used a rather low value of diffusivity.
As seen in Figure~\ref{diff}, \citet{MD14} and \citet{MT16} followed these authors in using a diffusivity
which was, within the body of the convection zone, about one order of magnitude smaller
than what we are using. In our case the diffusivity $\eta_{\rm mid}$ in the convection zone
is $5\times 10^{11}$ cm$^2$s$^{-1}$, whereas \citet{MD14} and \citet{MT16} use $5\times10^{10}$ cm$^2$s$^{-1}$.
Such a lower value of diffusivity would make the diffusion time across
the convection zone of the order of 70 years and the advection by the meridional circulation
would clearly be the dominant process for the transport of the poloidal field. \citet{MD14}
presented a self-excited dynamo solution for this situation.  For the value of diffusivity
we are using, the diffusion across the convection zone is the primary process for bringing
the poloidal field from the solar surface to the tachocline.  We believe that we are the first
to obtain a self-excited 3D kinematic dynamo solution for this case, which we contend is closer to reality.

We now discuss how the source term $S(\theta,\phi,t)$ in (1) is specified with the help
of the SpotMaker algorithm to treat the BL process.
This algorithm is mainly a 3D generalization of the Durney's double ring algorithm 
\citep{Durney95,Durney97}. In this algorithm, two suitable opposite-polarity spots are placed 
on the surface of the Sun in response to the dynamo-generated field near the base of the 
convection zone and then they are allowed to decay in the presence of mean flows (meridional 
circulation and differential rotation) and diffusivity. The first aim of this algorithm 
is to find out the suitable position for these spots to be placed on the surface. To do so, 
we calculate the mean toroidal flux $\bar{B}(\theta,\phi)$ near the bottom of the 
convection zone averaged over the tachocline thickness\citep{MT16} and find out where this 
field is crossing the threshold value $B_t$. It is believed that if magnetic fields near 
the bottom of the convection zone are stronger than the threshold value $B_t$ then they become 
magnetically buoyant and create the bipolar sunspots on the surface \citep{Parker75}. 
So the latitude and longitude of the spot pair is chosen randomly from all grid points
where the mean toroidal flux exceeds $B_t$, subject to a mask that suppresses spots at high latitudes. When 
we are able to find the $\theta_s$ and $\phi_s$ where the dynamo-generated toroidal field 
is more than the threshold value $B_t$, we put two spots on the surface at that position. Once 
the position of the bipolar sunspots is decided, the next step is to specify the magnetic field
there, by putting some tilt angle between the two sunspots according to Joy's law.
For that we use the polynomial profile as given in \citet{MD14} and 
for tilt angle we follow the procedure given in \citet{SK12}. We choose the tilt angle to be 
$\delta = 32.1^{\circ} \cos\theta$. We do not want to put these spots at each time step of our simulation. 
There are always certain time differences between the appearances of different sunspot groups. 
So we have used a time delay probability density function which allows us to put 
successive sunspot pairs having a random time delay between their appearances
\citep{MT16}. Another thing we should mention here is that, as seen in the observed butterfly
diagram, sunspots are found mostly on the lower latitudes and in our model we artificially
suppress the sunspot formation at higher latitudes using some masking function as given
in equation (3) of the \citet{MD14}. The flux content in the spots and the strength of the radial field 
are chosen based on the dynamo-generated field $\bar{B}$ and the observed strength of the 
sunspots as given below.
\begin{equation}
\label{flux}
\Phi = 2\alpha_{spot}\frac{|{\hat{B}(\theta_s,\phi_s,t)}|}{B_q}\frac{10^{23}}{1 + (\hat{B}(\theta,\phi)/B_q)^2} \rm{Mx}
\end{equation} 
where $\hat{B} = g(\theta)\bar{B}$ is the toroidal field after using the masking function $g(\theta)$ 
to suppress sunspots at high latitude and $B_q$ is the quenching field strength.
Here, $\alpha_{\rm spot}$ is the parameter which determines whether the dynamo will be sub-critical or 
super-critical. Our ultimate aim would be to make the dynamo work with
$\alpha_{spot} =1$ so that the flux in a particular BMR will have a value of $10^{23}$ \rm{Mx} as 
observed in case of the subsurface field strength equivalent to the quenching field strength.
But if the subsurface field at the bottom of the convection zone is not close to the quenching field, 
then we have to increase the value of $\alpha_{\rm spot}$ in order to get a working dynamo with bigger 
spots. In case of the diffusion-dominated dynamo, we are able to get a working dynamo with $\alpha_{\rm spot}=100$. 
While creating sunspot pairs by the SpotMaker algorithm, once the total flux is fixed by (\ref{flux}), we
have the freedom of selecting either the magnetic field strength or the size. We choose the
magnetic field strength inside the sunspots to be 3000 G, which fixes the size.

Since we are solving magnetic fields in a 3D spherical shell, so we must have to specify the subsurface structure of the 
sunspots which are put on the surface using the SpotMaker algorithm. As it is argued by \citet{Longcope02} and
\citet{SM05} that the sunspots get quickly disconnected from the parent flux tube, we make a very simple potential 
field approximation for the sunspot fields (see Figure~2(a,b) of \citet{MT16}). We ensure that the 
radial field becomes zero at some penetration depth ($r =0.90R$)
and it is equal to the imposed sunspot field at the surface ($r = R$). For the upper boundary condition, we
take the magnetic field to be radial at the solar surface. Throughout our simulation we use $N_r$ = $200$, $N_\theta$ = $256$ and $N_\phi$ = $512.$
All of the cases where we show the field lines above the solar surface ($r = R$) are the extrapolated fields using a free potential approximation.
\begin{figure*}[!htbp]
\centerline{\includegraphics[width=1.0\textwidth,clip=]{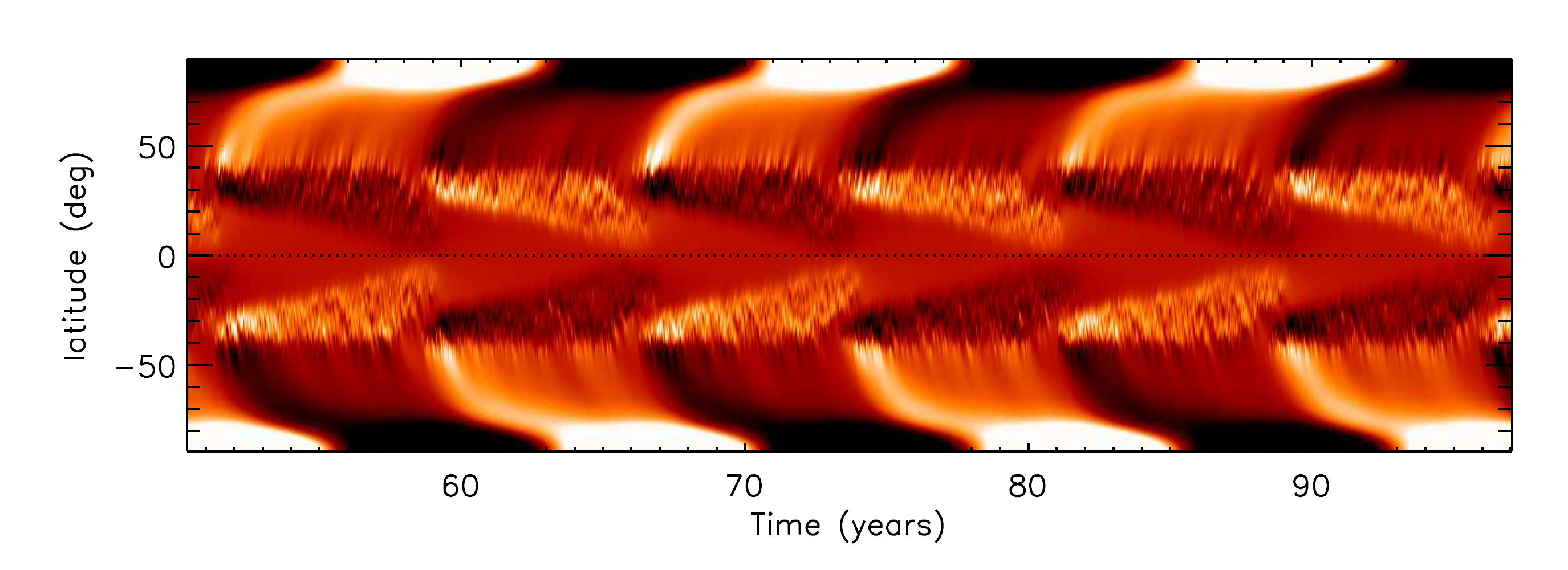}} 
\caption{Time-latitude plot of longitudinally averaged radial magnetic field on the surface $(r = R)$ of the Sun. Color table is set at $\pm10$ kG. }
\label{bfly}
\end{figure*}

\begin{figure*}[!htbp]
\centerline{\includegraphics[width=0.95\textwidth,clip=]{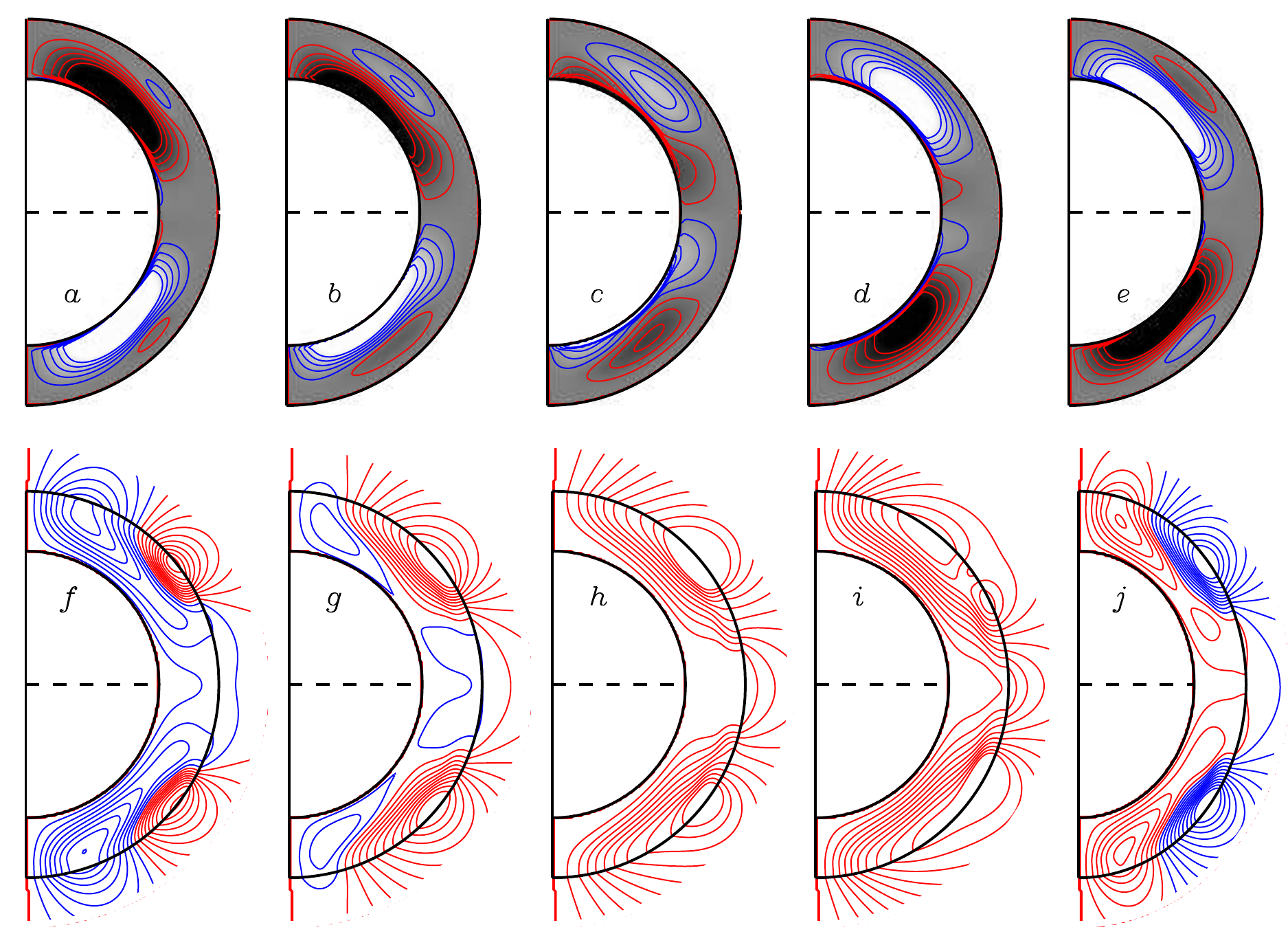}} 
\caption{Mean toroidal and poloidal field lines are shown for a particular solar cycle at five different 
time: t= 60.0 (a), (f), 62.0 (b), (g), 64.0 (c), (h), 66.0 (d), (i) and 68.0 years (e), (j). Frames (a)-(e) show mean 
toroidal fields with red and blue lines indicating eastward and westward field respectively. 
Filled color also represents the mean toroidal fields. Color table is set in this case at $\pm 500$ kG. 
Frames (f)-(g) represent poloidal magnetic potential with potential field extrapolation above the surface (upto $r = 1.25R$) where red 
and blue lines represent clockwise and anticlockwise directions. The maximum and minimum contour level is set corresponding to the poloidal field strength of $\pm49$ kG}
\label{fields_ref}
\end{figure*}

\section{Our reference model}

Now we present a self-excited solution from our reference model with parameters as prescribed
in the previous section.  To the best of our knowledge, this is the first self-excited
3D kinematic dynamo solution in which the diffusivity has been assumed sufficiently high to
make sure that the poloidal field is transported from the surface to the tachocline primarily
by diffusion.  The earlier results presented by \citet{MD14} and \citet{MT16} were obtained
with a diffusivity one order of magnitude smaller and the transport of the poloidal field was
due to the meridional circulation.

Figure~\ref{bfly} shows a butterfly diagram obtained by putting the longitude-averaged $B_r$ in a time-latitude
plot.  One clearly sees the butterfly diagram of sunspots at lower latitudes and the poleward
advection of the magnetic field by the meridional circulation at higher latitudes.  Superficially, this resembles
Figure~6(a) of \citet{MT16}, although our solution is for the diffusion-dominated case in contrast
to the solution of \citet{MT16} obtained for the case dominated by advection due to the meridional
circulation. The differences between the two cases become clear on looking at the distribution
of the magnetic field. Figure~\ref{fields_ref} shows the evolution of the toroidal and the poloidal fields during
a cycle. Comparing with Figure~8 of \citet{MT16}, we see some obvious differences.  In the solution
of \citet{MT16}, the oppositely directed toroidal fields on the two sides of the equator almost
pressed against each other.  Due to the low diffusivity, there would not be much diffusion of the
toroidal field even when two opposite bands are brought so close to each other.  In our model with
higher diffusivity, however, there would be more diffusion of the toroidal field across the equator, making
sure that the bands of concentrated opposite polarity are kept somewhat apart, as seen in Fig.~\ref{fields_ref}.

\begin{figure*}[!htbb]
\centerline{\includegraphics[width=1.0\textwidth,clip=]{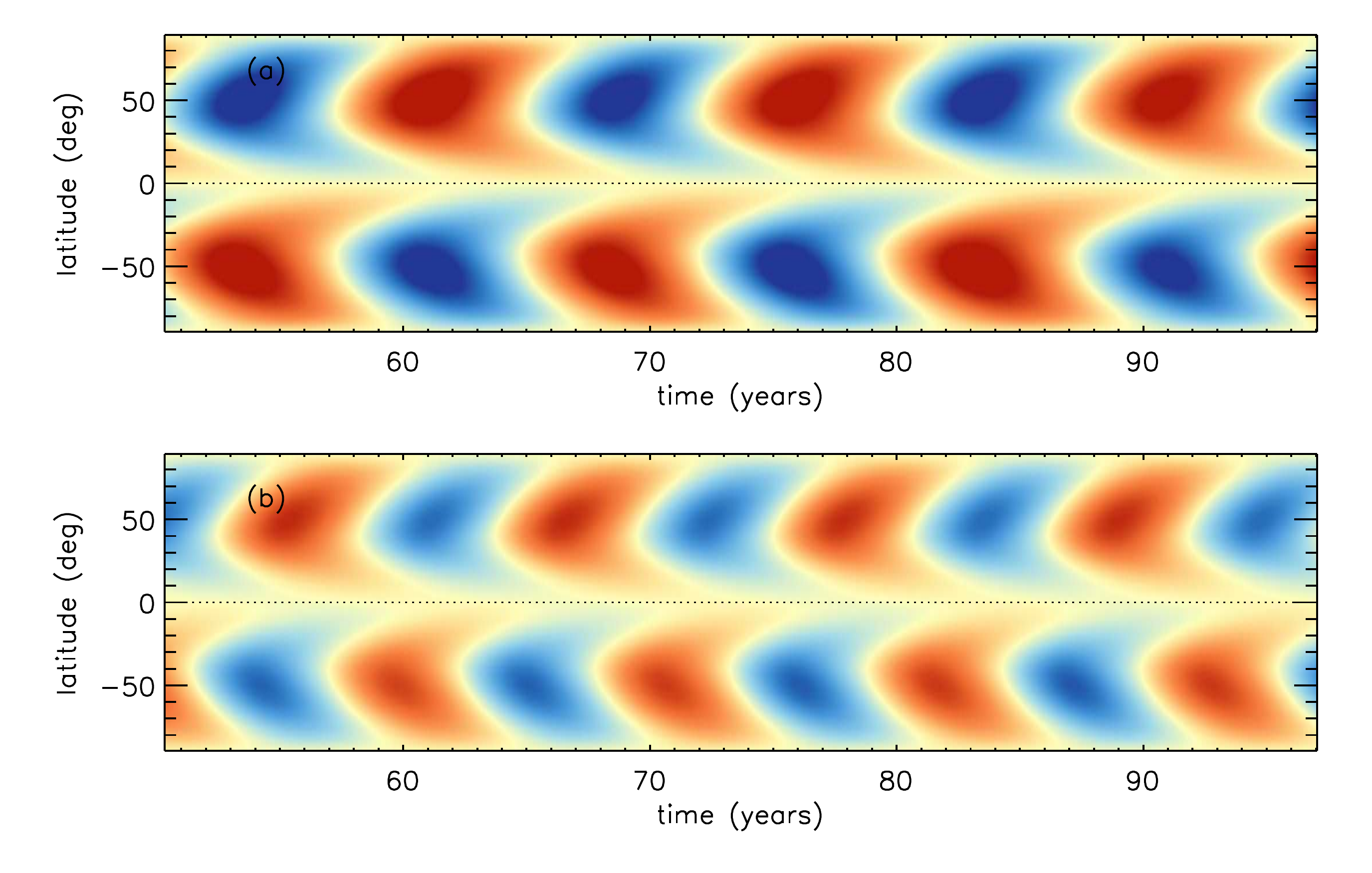}} 
\caption{Azimuthal averaged toroidal fields $B_{\phi}$ at the bottom of the convection zone $(r = 0.71R)$
for two different values of diffusivity $\eta_{\rm mid}$ at the convection zone
(a) $5\times10^{11}$ cm$^2$ s$^{-1}$ and (b) $7\times10^{11}$ cm$^2$ s$^{-1}$. Color table is set at $\pm  400$ kG.}
\label{bfly2}
\end{figure*}

The solar magnetic field is predominantly dipolar.  One requirement of a theoretical solar dynamo model is
that it should have dipolar parity. We have run our reference model for several cycles to ensure
that the dipolar parity persisted. One important question is, under what circumstances we
would expect dipolar parity.  This question has been studied thoroughly by
\citet{CNC04} and \citet{Hotta10} for the 2D kinematic dynamo model.  A full study
of this question requires running the code for many different combinations of parameters and
running it for a large number of cycles for each such combination. This would require a huge
amount of computer time for the 3D model.  Because of the limited computer time available to
us, we have not been able to study this question exhaustively.  However, we have made a limited 
number of runs to explore the issue of parity a little bit.  The best way to look at the issue
of parity is to make a butterfly diagram of longitudinally averaged $B_{\phi}$ at the bottom of the convection zone, as
done in Figure~7(a) of \citet{CNC04}. In Figure~\ref{bfly2}(a) we show such a plot for
our reference model, whereas Figure~\ref{bfly2}(b) shows a similar plot for the case in which the value
of diffusivity within the convection zone has been changed from  $5 \times 10^{11}$
cm$^2$ s$^{-1}$ to  $7 \times 10^{11}$ cm$^2$ s$^{-1}$ while keeping all the other
parameters exactly the same as in our reference model. We clearly see in Figure~\ref{bfly2}(b) that the 
nature of the solution is changing from a dipolar parity to a quadrupolar parity.

The surprising fact is that we now seem to get a result which is the opposite of what
\citet{CNC04} and \citet{Hotta10} obtained for the 2D kinematic dynamo model. These
authors found that the dipolar parity is preferred on increasing the diffusivity, whereas we
now are finding the opposite of that. Let us look at the physics of the problem.  In a dipolar
mode, the poloidal magnetic field lines connect across the equator, whereas the toroidal field
on the two sides of the equator has to be directed oppositely.  In order for this to happen, we need
diffusivity to have a big effect on the poloidal field, but not much effect on the toroidal field.
In the model of \citet{CNC04}, any strong toroidal field within the 
convection zone was removed by magnetic buoyancy and the toroidal field at the bottom of the
convection zone was also depleted continuously to account for flux loss due to magnetic buoyancy.
As a result, the toroidal magnetic field near the equator was naturally weak and the effect of
diffusion was more important on the poloidal field than on the toroidal field.  This ensured
that higher diffusivity favored the dipolar mode.  In the present calculation, the situation
is rather different. The strong parts of the toroidal field are now allowed to hover in the 
middle of the convection zone and near the equator.  On increasing diffusivity, the quadrupolar
mode in which the toroidal field on the two sides of the equator has the same sign is favored.
Here, we should mention that the magnetic pumping can play a very important role to promote
dipolar parity \citep{Guerrero08}. Another point to note is that \citet{CNC04} used a lower
diffusivity of the toroidal field compared to the poloidal field, to account for the quenching
of turbulent diffusion due to the stronger toroidal magnetic field.  This could be done in a
2D mean field model in which the evolution equations for the toroidal and poloidal fields neatly
separate out, and one could use different values of diffusivity in the two equations. Since it
is not possible to do this in a 3D non-axisymmetric model in which the equations for the toroidal and 
poloidal components do not split in this way, we have used a single diffusivity. It is possible
that the weaker diffusivity of the toroidal field in \citet{CNC04} helped in producing a dipolar
parity by allowing toroidal fields of opposite sign to exist on the two sides of the equator
more easily. One way of capturing the physics of this in a 3D non-axisymmetric model may be
to include a quenching of turbulent diffusivity in the regions of strong magnetic field.  We
plan to explore the effect of this in future.

With these two opposite results at hand, one crucial question is: which of the two results is closer
to reality?  Although we cannot assert this with confidence at this stage, we believe
that the 2D kinematic dynamo result that the dipolar parity is preferred on increasing diffusivity
is the more appropriate result.  Although in this paper we are taking account of the 3D nature of
magnetic buoyancy and, in that sense, treating magnetic buoyancy more realistically, we still
have not taken account of flux depletion from the convection zone and its bottom in an appropriate
way. This is probably one important reason why our results are not matching with 
the results of previous 2D models \citep{CNC04,Hotta10}.
We are right now exploring possible schemes to take account of the flux depletion due to magnetic buoyancy in a
realistic way.  We believe that this flux depletion is quite important in the solar dynamo.
\citet{CH16} found that the Waldmeier effect cannot be reproduced from a theoretical
dynamo model unless the flux depletion is taken into account. We have a future plan of incorporating
flux loss due to magnetic buoyancy in a realistic way in our 3D kinematic dynamo model and then
studying the parity issue more carefully.

Since we are interested in a dynamo solution which has dipolar parity, we have converged on
the reference solution presented here.  If we decrease diffusivity, then we are led to the case
where the meridional circulation provides the main transport mechanism for the poloidal field.
On the other hand, if we increase diffusivity, then we obtain the quadrupolar mode.  This is what
has led us to choose the value $5 \times 10^{11}$ cm$^2$ s$^{-1}$ for diffusivity inside the
convection zone. 

\section{The build-up of the polar field}

After constructing the self-excited dynamo model, we now study 
how individual sunspot pairs contribute to the building up of
the polar field and address the question whether our understanding gained from
this study necessitates the revision of some insights we have from surface flux
transport (SFT) models. For this study, we shall put individual sunspot pairs on
the solar surface by hand and look at the evolution of the magnetic field.  In
other words, we shall now not try to construct self-excited periodic solutions,
although we shall keep using the same values of different parameters that we had
used for constructing the self-excited periodic solution.

We start our simulation by putting a single pair of bipolar sunspots in the northern hemisphere at 
different emergence angles $\lambda_{\rm emg}$ and let it evolve under the axisymmetric mean 
flows and diffusion to see the development of the polar field. We have chosen magnetic flux 
of $1\times10^{22}$ \rm{Mx} in each spot and its radius is taken to be 21.71 Mm (somewhat
larger than actual sunspot radii, to make the results of the simulation more clearly visible)
throughout our simulations. In the next set of our simulations, we shall put two pairs of sunspots  
symmetrically in the two hemispheres, which have the same amount of flux and radius 
as in the case of the single pair, to see the effects of cross-equator diffusion of magnetic flux. 

\subsection{Polar field from one sunspot pair}

\begin{figure*}[!htbp]
\centerline{\includegraphics[width=1.0\textwidth,clip=]{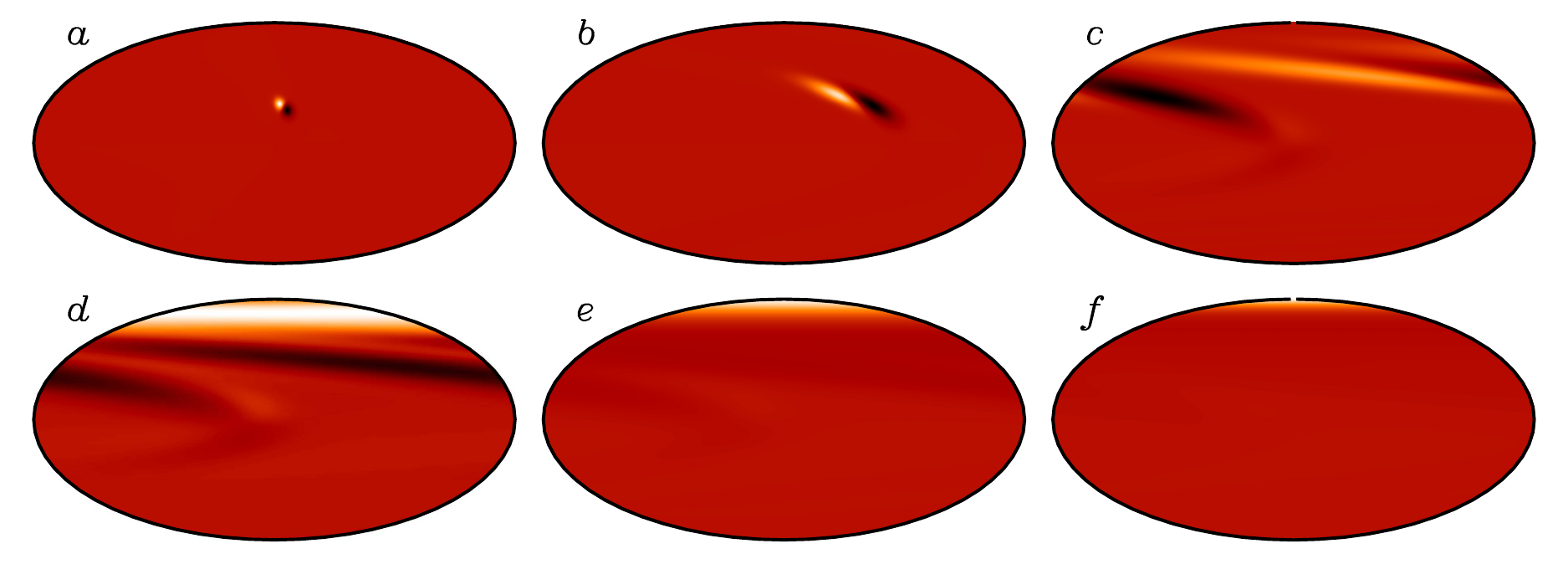}}
\caption{Time evolution of radial fields on the surface of the sun with a single pair in the 
northern hemisphere for (a) 0.025 yr, (b) 0.25 yr, (c) 1.02 yr, (d) 2.03 yr, (e) 3.05 
yr and (f) 4.06 yr. Here white color shows the outward-going radial field and black color 
represents inward-going radial field. The color scale is set at $\pm$maximum values of the magnetic fields for 
each case. For example $\pm4.66$ G is the color scale for (a) and $\pm0.10$ G is the color scale for (f).}
\label{sfield_1}
\end{figure*}

\begin{figure*}[!htbp]
\centerline{\includegraphics[width=1.0\textwidth,clip=]{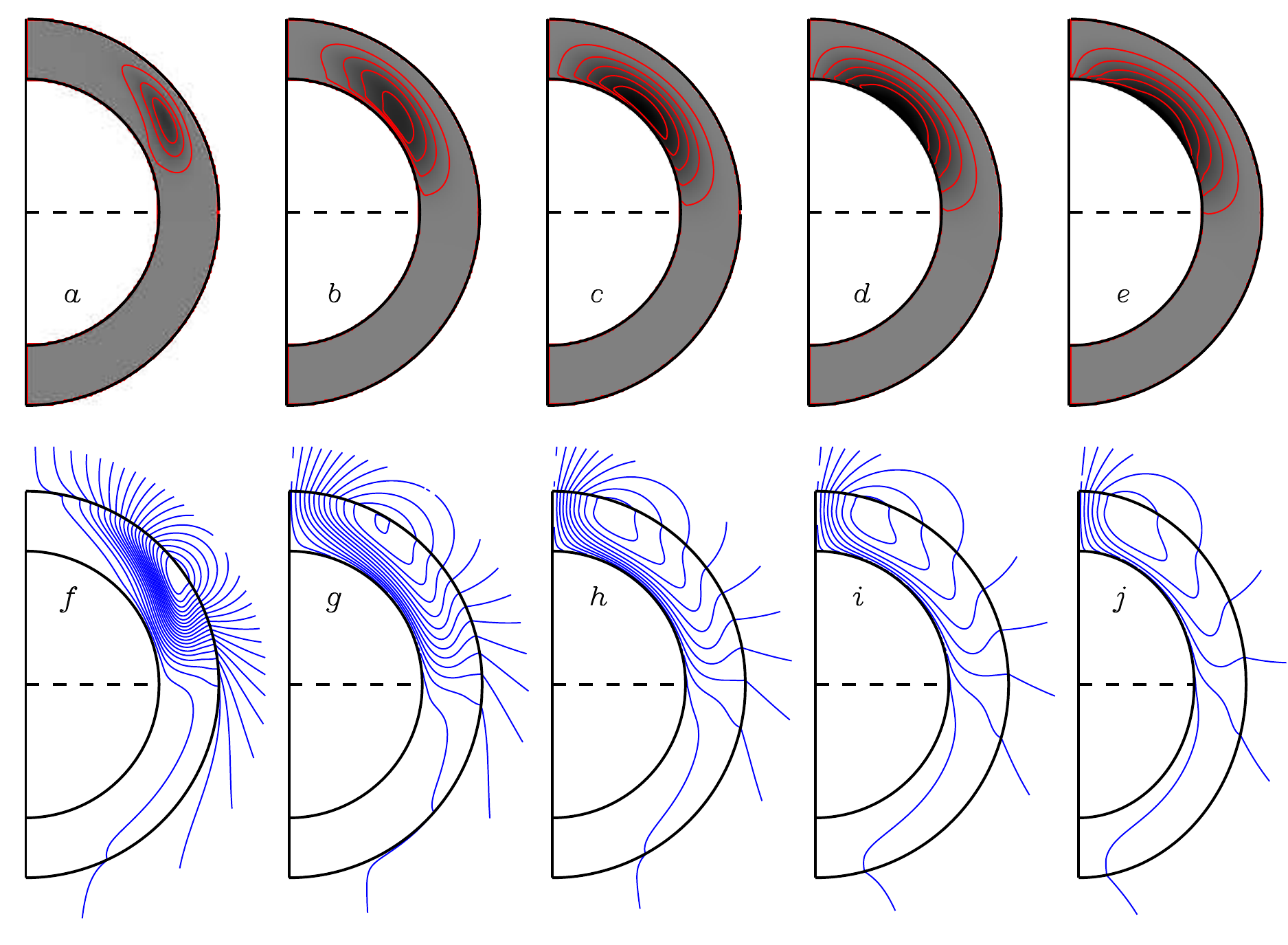}} 
\caption{Axisymmetric toroidal field lines (a)-(e) and axisymmetric poloidal field lines (f)-(j) are 
shown for 5 different times. Time spans are (a), (f) = 1.02 yr, (b), (g)= 3.05 yr, (c), (h) = 5.08 yr, 
(d), (i) = 7.11 yr and (e), (j) = 9.15 yr. Frames (a)-(e) represent $<B_{\phi}>$ (azimuthal averaged) 
with red and blue indicating eastward and westward fields respectively. Filled contour also
represents the mean toroidal fields. Here color scale is set 
at $\pm1.5$ G. Frames (f)-(j) represent the square root of poloidal magnetic potential with potential field extrapolation
above the surface (up to $r = 1.25R$) and blue color contours denote the clockwise direction of the field. 
Maximum and minimum contour levels are set corresponding to potential field strength of $\pm0.3$ G respectively.}
\label{mfield_1}
\end{figure*}

We use the SpotMaker algorithm to put one sunspot pair at latitude $20^{\circ}$ with tilt
angle $40^{\circ}$. Then, we allow our code to evolve the magnetic field from this sunspot
pair leading to the build-up of the polar field. Figure~\ref{sfield_1} shows snapshots of $B_r$ on
the solar surface at different times during the evolution process. This figure can
be compared with Figure~6 of \citet{YM13}. Although \citet{YM13} assumed the sunspot
pair to be initially connected to the toroidal flux system at the bottom of the
convection zone, eventually this connection would be disrupted and the evolution
of the magnetic field on the surface due to the sunspot pair in the northern hemisphere
appears to be very similar to the evolution that
we get by assuming a disconnection from the very beginning. The following
sunspot at the higher latitude has the positive polarity and we clearly see that
this positive polarity is preferentially transported to the higher latitudes. This
positive polarity region gets stretched by the differential rotation into a belt going
around the polar axis. When this belt reaches sufficiently high latitude, we see
that it is followed by a belt of negative polarity coming from the leading sunspot
which was taken at a lower latitude. The meridional circulation takes about 3
years to bring the flux of $B_r$ to create a positive patch on the pole surrounded a
ring of negative polarity. The formation process of the negative polarity ring is clearly visible in
Figure~\ref{sfield_1}(d), but at later times it becomes weaker due to the action of diffusion
and is not clearly visible. Since the meridional circulation sinks downward at the
polar region, eventually both the positive and negative polarity magnetic fields
are advected simultaneously below the surface. This becomes clear from the field
line plots shown in Figure \ref{mfield_1}. At certain instants of time, we have averaged $B_r$ and
$B_{\theta}$ over the azimuthal direction $\phi$ to obtain the field lines.

It may be noted that the color scale for each plot in Figure~\ref{sfield_1}
is set at $\pm$ maximum values of the magnetic field in each case.  This was necessary
because the magnetic field becomes weak with time. Had we used the color scale of 
Figure~\ref{sfield_1} (a) for all cases shown in Figure~\ref{sfield_1}, then the magnetic
field would be completely invisible for the plots at later times.  Though the magnetic
fields in the sunspot pair remain concentrated for a shorter time than what one may suspect
from a casual look at Figure~\ref{sfield_1}, the sunspots in our simulations are
nevertheless live longer than real sunspots.  This is expected because we have assumed
the sizes of sunspots in our calculation to be larger than real sunspots.  If sunspots 
decay by the action of turbulent diffusion, then a simple application of 
the diffusion equation suggests that 
the lifetime should go as the square of the size.  A hypothetical sunspot 5 times larger
than a real sunspot should live 25 times longer than a real sunspot.

It should be kept in mind that $\int {\bf B}. d{\bf S}$ integrated over the whole solar surface has
to be zero at any time (since $\nabla. {\bf B} = 0$). This means that, during any time
interval, equal amounts of positive and negative magnetic fluxes have to disappear below 
the surface due to the subduction process. As a result, we see in Figure~\ref{sfield_1} that
the white patch at the pole (representing positive flux) remains there till all fluxes 
disappear and is not replaced by the poleward migrating dark ring (representing negative
flux, which gets subducted along with the positive flux).  In the real Sun 
undergoing successive cycles, the polar field reverses only when fluxes
of the following sunspots from the next cycle reach the pole. 
This subduction process has been seen before in axisymmetric
Babcock-Leighton / Flux-Transport dynamo models but has not been well studied within
the context of SFT models (though see \citet{Cameron12,YM13}).  
It relies on the upward advection of poloidal flux near the equator, which
leads to the emergence of oppositely-signed radial field, as shown in Figure~\ref{mfield_1}(f-j). 
This changes the net radial flux through the surface in each hemisphere
and eats away at the polar field as it is advected poleward.  
Without this low-latitude emergence, the subduction of poloidal flux
at the poles could not change the net flux through the outer surface.


In the SFT model also, a tilted sunspot pair gives rise to a polar field with the
polarity of the following sunspot surrounded by a belt of the opposite polarity.
However, since the low-latitude emergence and subsequent subduction of the mean
poloidal field is not included in the model, the net flux through each hemisphere
can only change by means of cross-equatorial transport and diffusion.
In a model of the solar magnetic field dynamics with
realistic values of various parameters, usually the diffusion time for neutralizing
the opposite magnetic polarities turns out to
be much longer than the time for their disappearance due to low-latitude emergence and subduction by the meridional circulation. 
\begin{figure*}[!htbp]
\centerline{\includegraphics[width=1.0\textwidth,clip=]{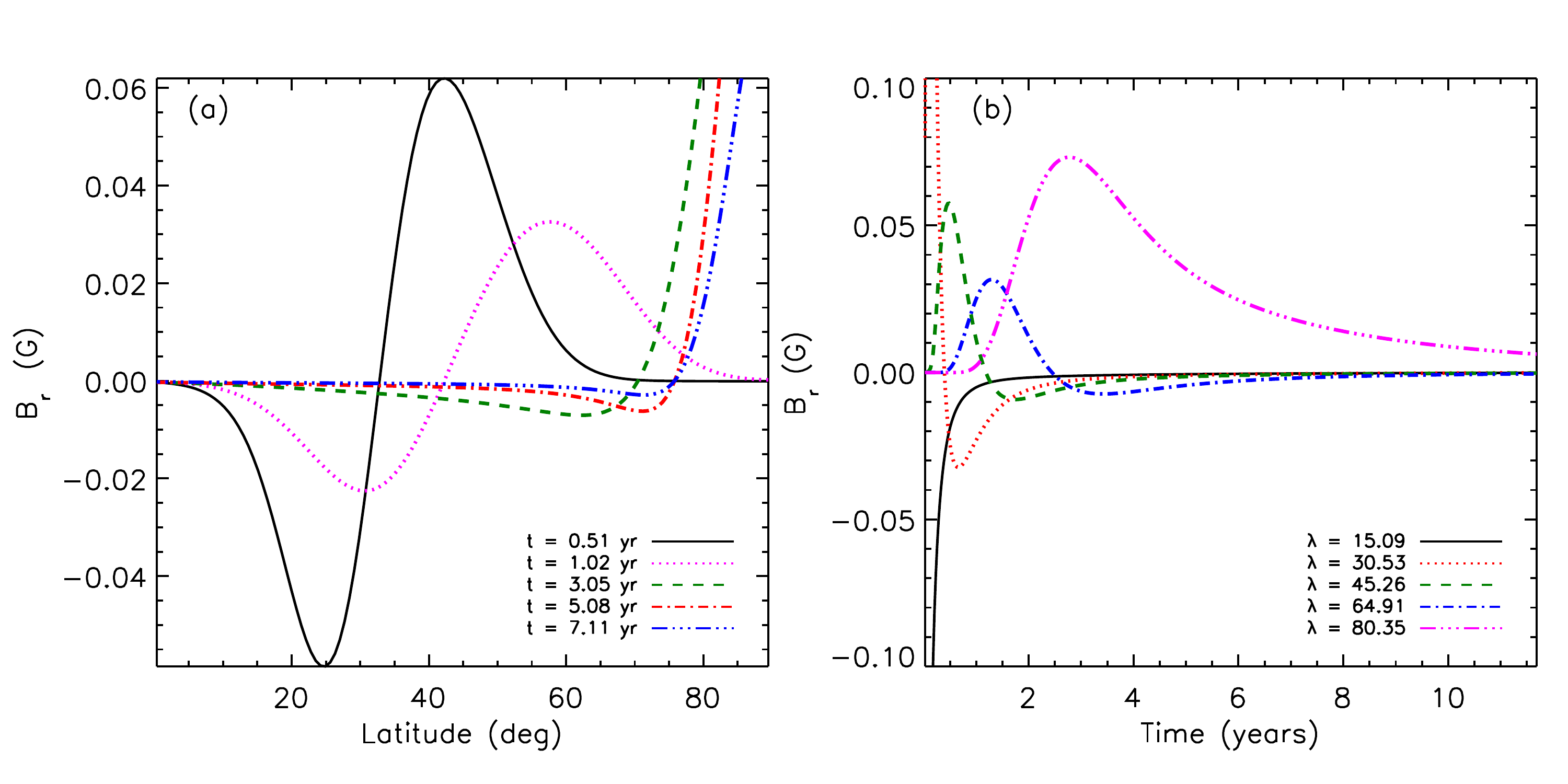}} 
\caption{(a) Behavior of radial field with latitude is plotted for different times. Radial field 
just after the emergence of the sunspot (at $20^{\circ}$ latitude) are shown in black line at 
time t = 0.51 yr. Magenta dotted, green dashed, red dash dotted, and blue long dash dotted lines represent variation of 
radial magnetic field with latitude at time 1.02 yr, 3.05 yr, 5.08 yr and 7.11 yr respectively. (b) Time variations 
of radial magnetic field for different latitudes are plotted. Solid black, red dotted, green dash dotted, blue dash dotted 
and magenta long dash dotted lines are for latitude $15^{\circ}.09$, $30^{\circ}.53$, $45^{\circ}.26$, $64^{\circ}.91$ 
and $80^{\circ}.35$ respectively. All units of magnetic fields are given in Gauss.}
\label{rfield_1}
\end{figure*}
\begin{figure*}[!htbp]
\centerline{\includegraphics[width=1.0\textwidth,clip=]{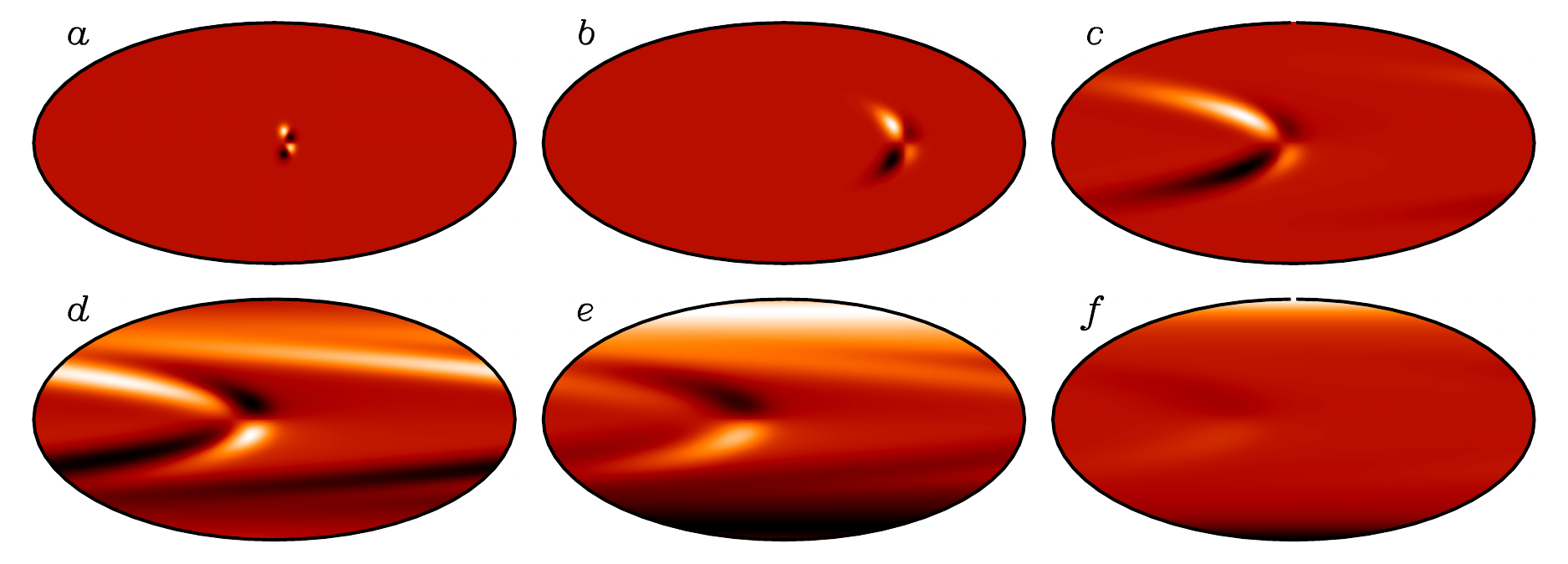}} 
\caption{Same as Figure~\ref{sfield_1} but for sunspot emergence in two hemispheres at $\pm5^{\circ}$ latitudes.
In this figure also color scale is set at $\pm$ maximum value of the magnetic fields for 
each case.}
\label{sfield_2}
\end{figure*}
As we see in Figure \ref{mfield_1}, the magnetic fields tend to sink below the surface 
while they diffuse and the disappearance of the magnetic fields at the surface
takes place in a time scale shorter than the diffusion time scale. We thus see that
the evolution of the polar field in our 3D model is qualitatively different from
what it is in the SFT model.

Figure~\ref{mfield_1} also shows the toroidal field generated in the convection zone.
Since the poloidal field has not yet reached the tachocline
to be acted upon by the radial differential rotation there, it may be worthwhile
to comment how the toroidal field is generated. As soon as we put a sunspot pair
on the surface by the SpotMaker algorithm, some toroidal field arises below the
surface at once because the magnetic loop connecting the two sunspots below the surface
would have a toroidal component.  Additionally, more toroidal field is
produced by the latitudinal differential rotation within the convection
zone.  It has been known that the latitudinal differential rotation can play an
important role in generating the toroidal field \citep{Guerrero07}.
In reality, any strong magnetic field generated within the convection
zone is expected to be quickly removed by magnetic buoyancy which is particularly
effective within the convection zone \citep{Parker75, Moreno83}. Since we
do not allow magnetic buoyancy to remove the toroidal field in the present version
of the code, the toroidal field remains where it is created. However, it may be 
noted that some fully dynamical simulations suggest persistent rings of toroidal
flux within the convection zone \citep{Brown10}.

Finally, Figure~\ref{rfield_1}(a) shows $B_r$ (averaged over $\phi$) as a function of latitude for
different times, whereas Figure~\ref{rfield_1}(b) shows $B_r$ as a function of time at different
latitudes.  A careful scrutiny of Figure~\ref{rfield_1}(a) makes it clear that the poleward meridional
circulation transports the magnetic flux to higher latitudes with time.  After about
3 years, the polar field starts building up.  It is clear that the polar field becomes
much stronger than the fields at mid-latitudes.  This is purely a geometrical effect.
Since magnetic flux from different longitudes is brought by the meridional circulation
to the pole where it converges, it is natural that the magnetic field becomes stronger
at the pole.  It is also to be noted that we only have the polar field with polarity
corresponding to the polarity of the following sunspot at the higher latitude (positive
in the present case).  Turning to Figure~\ref{rfield_1}(b) now, we first look at the plots corresponding
to the mid-latitudes ($\approx 30^{\circ} - 65^{\circ}$).  At a mid-latitude, first the magnetic field corresponding to
the polarity of the following sunspot (positive in the present case) is brought by
the meridional circulation, followed by the magnetic field with opposite polarity
from the leading sunspot (negative in the present case) a little bit later. This is
seen in all the mid-latitude plots in Figure~\ref{rfield_1}(b).  But we should pay a special attention
to the plots for latitudes $15^{\circ}$ and $80^{\circ}$.  At the latitude of  $15^{\circ}$,
the positive magnetic field from the following sunspot is never seen, because the
following sunspot appeared at a higher latitude and the meridional circulation transported
the flux from its decay toward the pole. On the other hand, at the latitude of  $80^{\circ}$,
we see only the positive magnetic field which has been brought there from the following
sunspot.  The negative magnetic field from the leading sunspot forms a negative
polarity belt around the pole, as we have already seen, and then it sinks below
the surface, so the negative magnetic field is never seen at sufficiently high latitudes.
Also, note that, although the peak value of the positive polarity field at $65^{\circ}$ is
less than that at $45^{\circ}$ (due to the action of diffusion while the magnetic field
is transported to higher latitudes), the positive polarity field again becomes strong
at $80^{\circ}$ due to the geometrical effect of converging flow bringing magnetic flux
from different longitudes.

\begin{figure*}[!htbp]
\centerline{\includegraphics[width=0.95\textwidth,clip=]{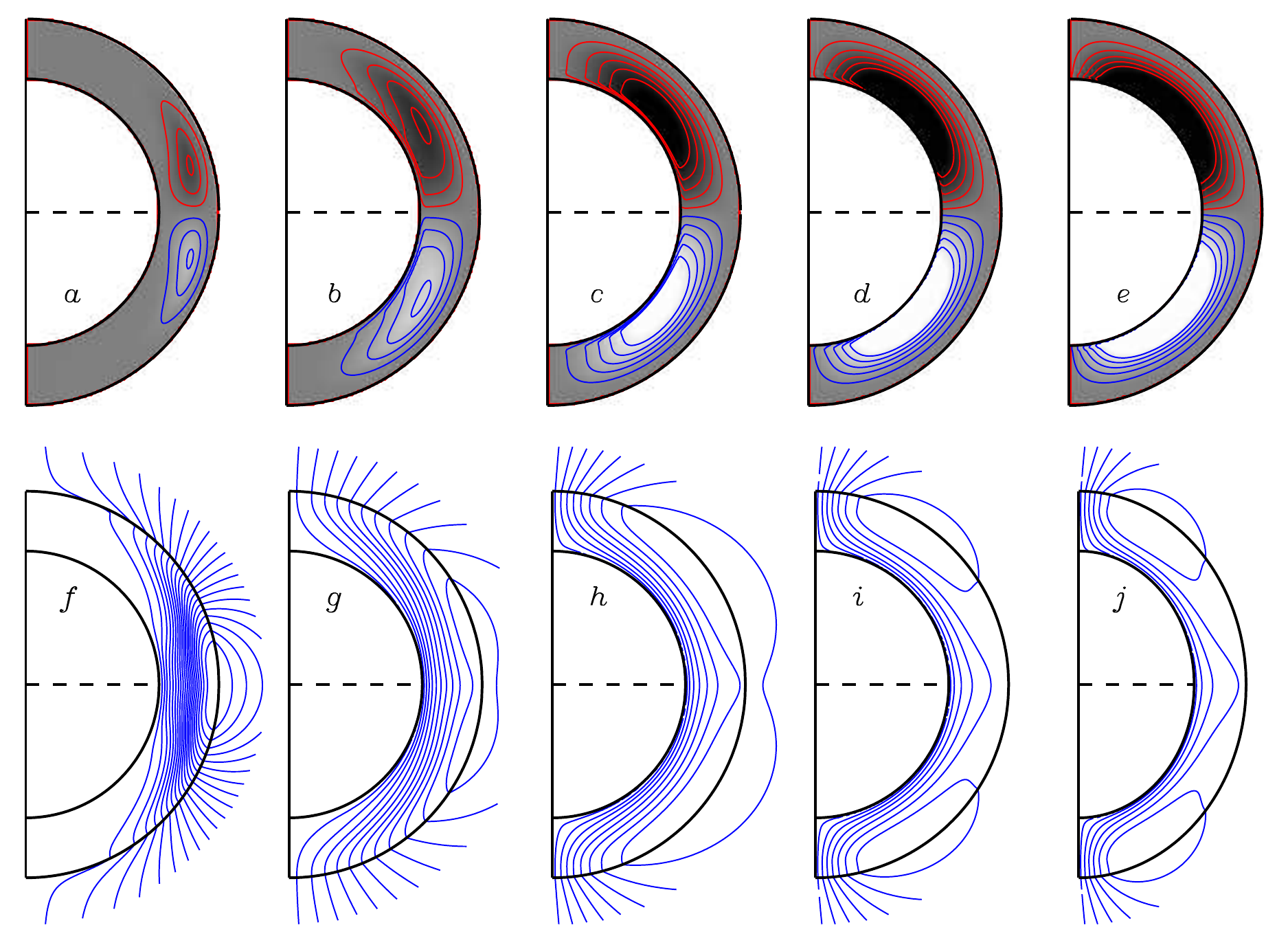}}
\caption{Same as Figure~\ref{mfield_1} but for two pairs at two hemispheres at $\pm5^{\circ}$ latitudes. 
Color scale for toroidal fields is set at $\pm 1.5 G$ and contour levels corresponding to the poloidal fields strengths of $\pm0.02$ G are set as maximum and minimum, respectively.}
\label{mfield_2}
\end{figure*}
\begin{figure*}[!htbp]
\centerline{\includegraphics[width=1.0\textwidth,clip=]{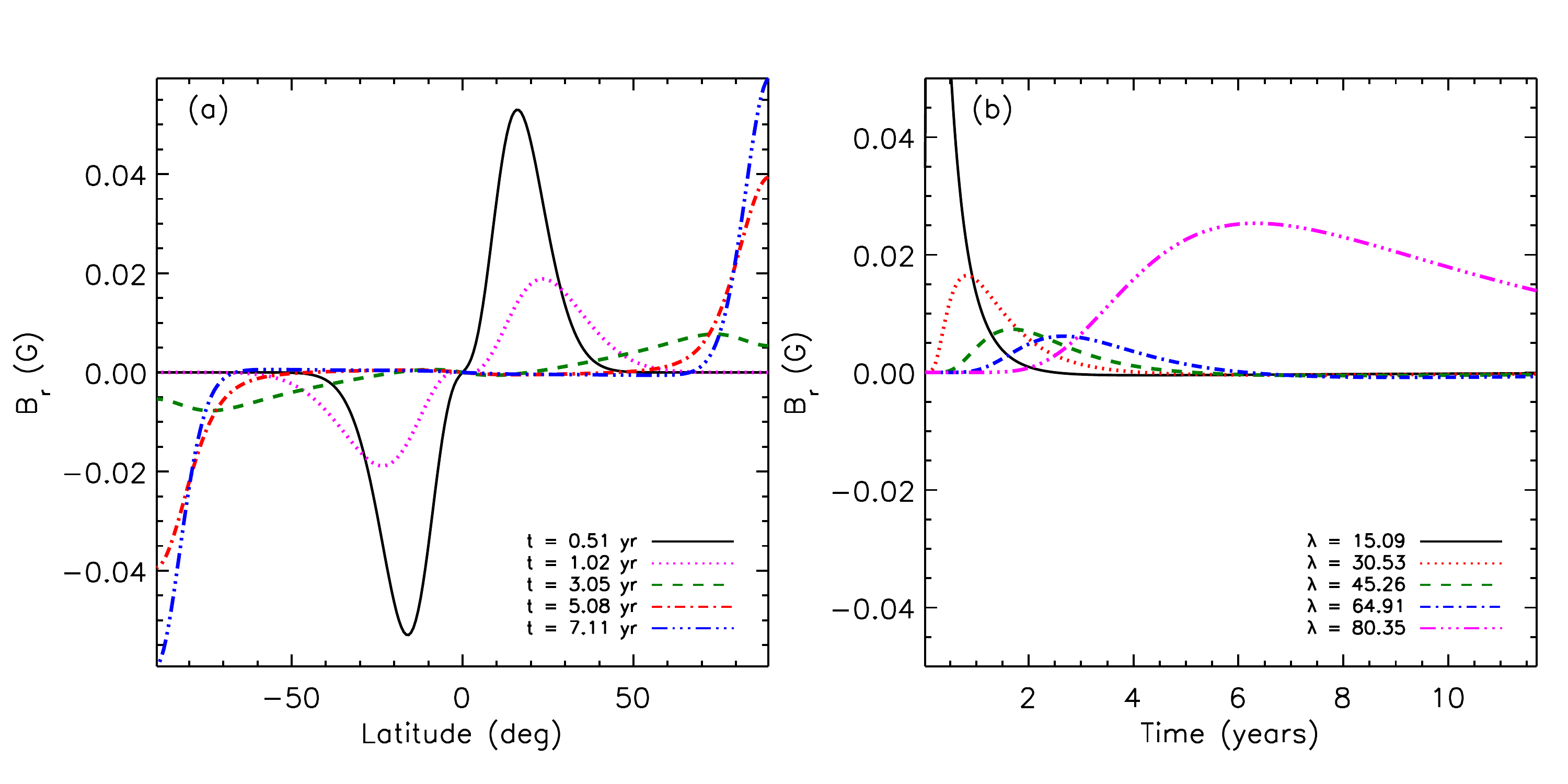}} 
\caption{Same as Figure~\ref{rfield_1} but with two pairs in two hemispheres at $\pm5^{\circ}$ latitudes.}
\label{rfield_2}
\end{figure*}

\subsection{Polar fields from two sunspot pairs in two hemispheres}

The results of the 3D model differ more dramatically from the results of the SFT model 
when we put two pairs of sunspots located symmetrically in the two hemispheres.  If the 
two pairs are sufficiently close to the equator, then magnetic fluxes of the two leading 
sunspots get canceled by diffusion across the equator. In the SFT model, only the fluxes
from the following polarities are advected to the two poles and we eventually get polar
patches which are not surrounded by rings of opposite polarity as we found in the case
of the single sunspot pair. When the outward spreading of magnetic field from the polar
patches by diffusion is eventually balanced
by the inward advection by the meridional circulation, we reach an asymptotic steady
state in the SFT model, with an asymptotic magnetic dipole which does not change with time.
This is seen in Figure~6 of \citet{Jiang14}.  As we shall discuss now, we get a completely
different result from the 3D model.

We see in Figure~\ref{sfield_2} that polar magnetic patches form with the polarity of
the succeeding sunspots. A careful look at this figure, shows some evidence of opposite 
polarity (i.e.\ opposite of what we see in the poles) at mid-latitudes 
even when we start from two sunspots placed symmetrically at sufficiently low latitudes in
both the hemispheres. The physics of what is happening becomes clear from the plot of
field lines shown in Figure~\ref{mfield_2}. After the fluxes from the leading sunspots near the equator
cancel, we see that initially we get poloidal field lines spanning both the hemispheres.
A look at the field line plots makes it clear that we shall have $B_r$ only at high latitudes
in the early stages of the evolution of the magnetic field. As the meridional circulation
drags the poloidal field toward the poles, we find that eventually the polar fields in the
two hemispheres get detached, as a result of which $B_r$ again appears at lower latitudes
having the opposite polarity of $B_r$ at high latitudes. This is purely a result of the
3D structure of the magnetic field and cannot happen in the SFT model. There would not be
a source for creating $B_r$ at low latitudes in the SFT model and such fields would never
appear in that model.  Because of the breakup of the poloidal field in the two hemispheres
and the appearance of $B_r$ with opposite polarity in the low latitudes, it is possible
for the poloidal magnetic field in the 3D model to be subducted below the surface as the
meridional circulation sinks downward in the polar regions. Thus, in contrast to the SFT
model in which polar fields have nothing to cancel them and therefore persist, the polar
field disappears after some time in the 3D model. 

Though this result is notable, it may be offset to some extent by efficient magnetic 
pumping.  Using a 2D (axisymmetric) model \citet{KC16} have shown that downward 
magnetic pumping due to strongly stratified convection in the solar surface layers can 
suppress the upward diffusion and advection of toroidal and poloidal fields.  This, 
in turn, can produce steady polar fields that might persist indefinitely.  We will 
investigate the role of magnetic pumping in future work.

Figure~\ref{rfield_2}(a) is similar to Figure~\ref{rfield_1}(a) except that latitudes now cover from 
$-90^{\circ}$ to $90^{\circ}$. In this figure, we clearly see that around 1 year, we had only 
positive $B_r$ in the northern hemisphere and negative $B_r$ in the southern hemisphere, 
but afterwards very weak $B_r$ having sign opposite to the sign at the high latitudes
developed at low latitudes. Figure~\ref{rfield_2}(b), which is similar to Figure~\ref{rfield_1}(b), shows 
that eventually $B_r$ disappears at the surface in this 3D model, exactly similar to what happens when we
put only one sunspot pair on the solar surface.

\begin{figure}[!htbp]
\centerline{\includegraphics[width=0.5\textwidth,clip=]{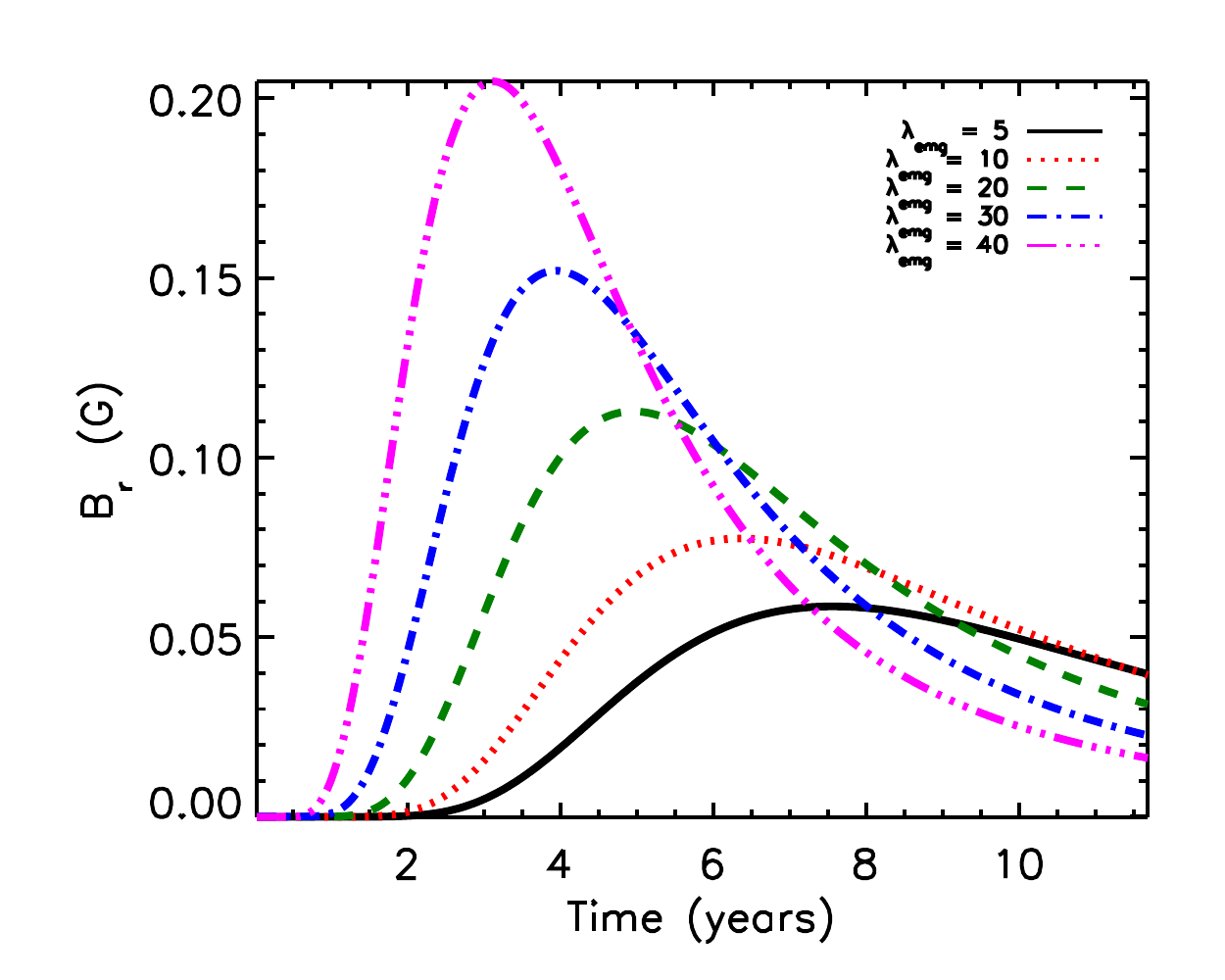}} 
\caption{Polar field evolution with time for different emergence angle $\lambda_{\rm emg}$ of sunspot pairs
in both the hemispheres. Black solid, red dotted, green dashed, blue dash dotted and
magenta long dash dotted lines represent the polar field for the sunspot emergence at $5^{\circ}$, 
$10^{\circ}$, $20^{\circ}$, $30^{\circ}$, and $40^{\circ}$ respectively. Magnetic field is in Gauss and time is given in years.}
\label{pfield_2}
\end{figure}

We carry on such calculations by putting two sunspot pairs symmetrically at different
latitudes in the two hemispheres. Figure~\ref{pfield_2} shows how the polar field evolves with time
for sunspot pairs placed at different latitudes.  When the sunspot pairs are placed at
high latitudes, the magnetic flux is brought to the poles without too much diffusion and
the polar field is stronger. Eventually, the polar field disappears in all the cases due to
emergence and subduction by the meridional circulation, as we have already discussed.
This figure can be compared with the left panel of Figure~6 of \citet{Jiang14}.  Such
a comparison makes the difference between the 3D model and SFT model completely clear.
In the SFT model, only if the sunspot pairs are put at sufficiently high latitudes so
that cross-equatorial diffusion is negligible, fluxes of both polarity are advected to
the polar regions and eventually the axial dipole moment becomes zero.  If the sunspot
pairs are put at low latitudes in the SFT model, only the fluxes from the following sunspots
reach the poles and give rise to an asymptotic axial dipole.  The situation is completely
different in the 3D model, although we see that the polar field persists for a longer time
when the initial sunspot pairs are put at lower latitudes. So, in that sense, sunspot
pairs appearing in lower latitudes are somewhat more effective in creating the polar
field even in the 3D model. This is in agreement with the claim of \citet{DasiEspuig10}
that we have a better correlation between the average tilt of a cycle and the strength of
the next cycle if more weight is given to sunspot pairs at low latitudes when computing
the average tilt. 

\begin{figure}[!htbp]
\centerline{\includegraphics[width=0.5\textwidth,clip=]{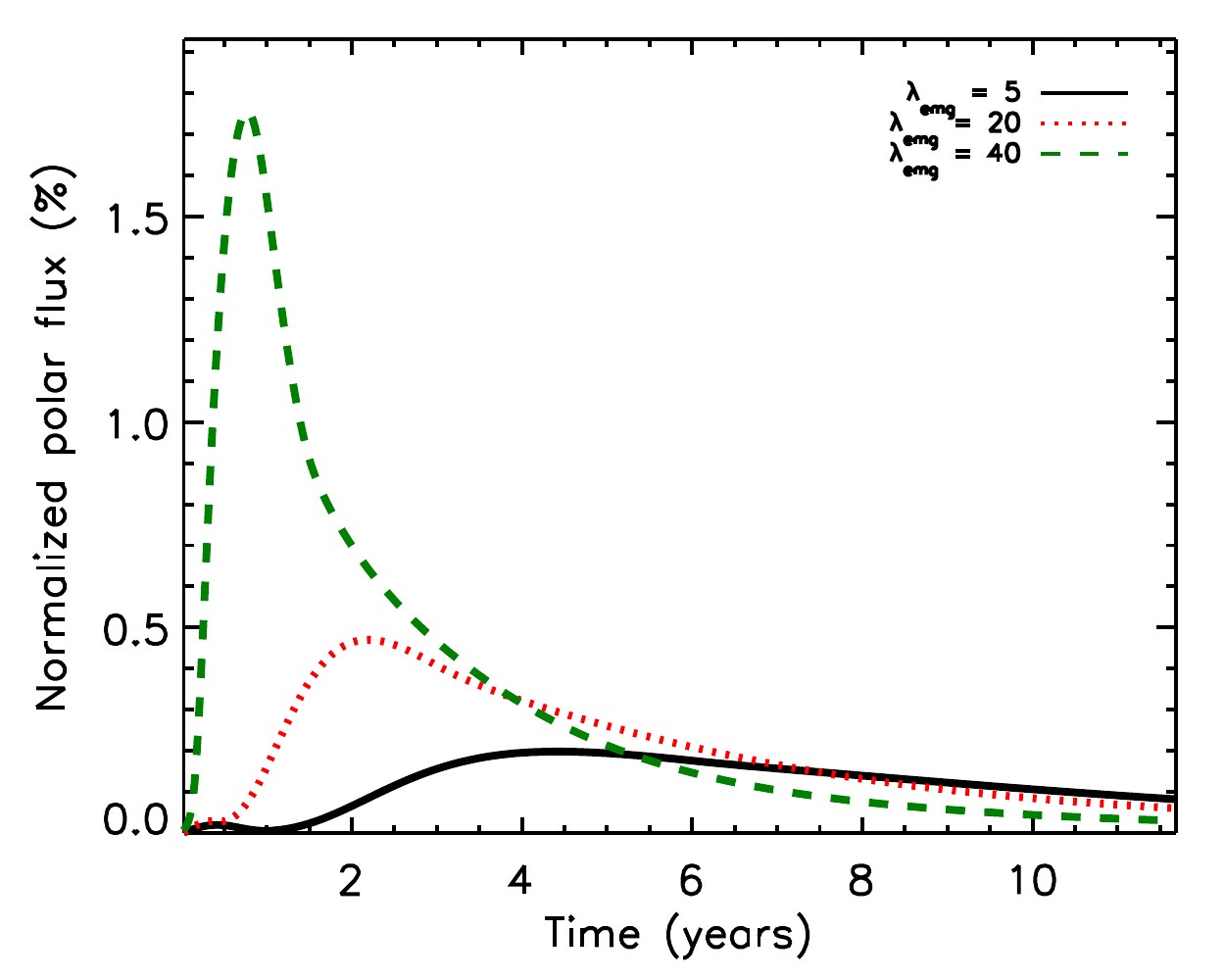}} 
\caption{Polar flux evolution with time for different emergence angle $\lambda_{emg}$ of sunspot pairs
in both the hemisphere. Black solid, red dotted, green dashed lines 
represent the percentage of normalized polar flux able to reach the pole for the sunspot emergence at $5^{\circ}$, 
$20^{\circ}$ and $40^{\circ}$ respectively.}
\label{pflux_2}
\end{figure}

\begin{figure*}[!htbp]
\centerline{\includegraphics[width=1.0\textwidth,clip=]{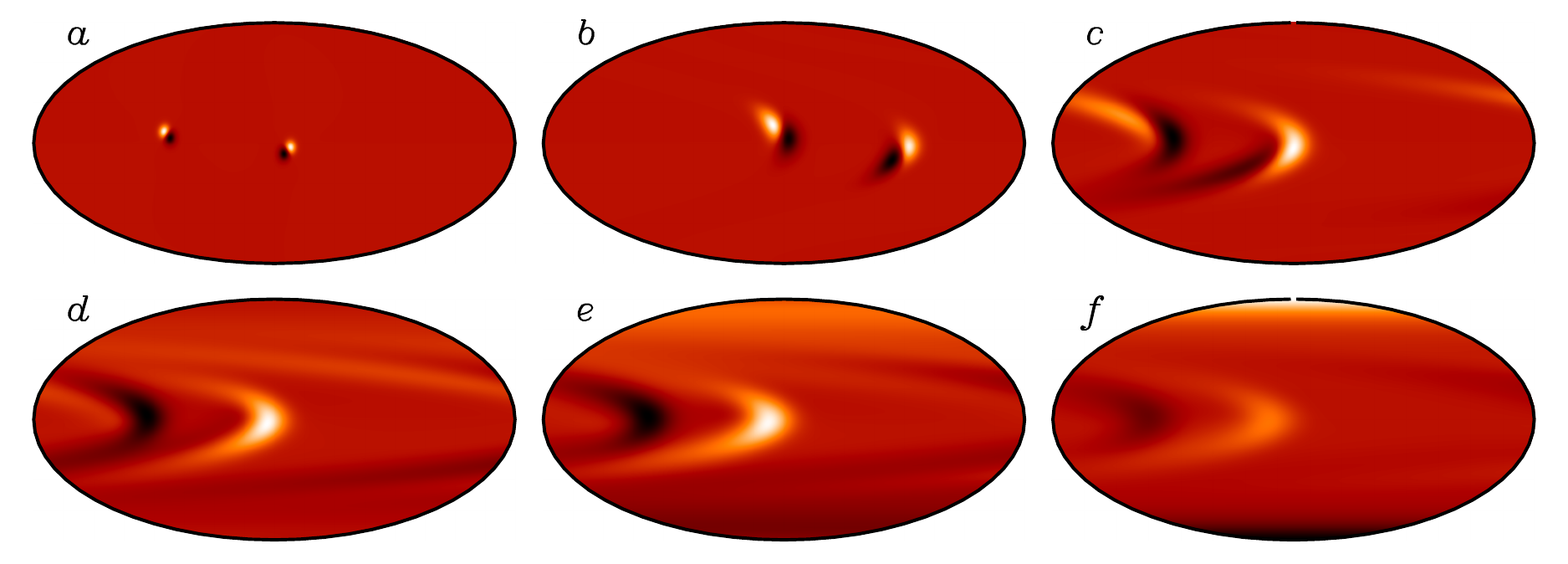}} 
\caption{Same as Figure~\ref{sfield_2} but sunspot pairs are placed at longitude $90^{\circ}$ on
northern hemisphere and at longitude $180^{\circ}$ on southern hemisphere.}
\label{sfield_3}
\end{figure*}

We have also calculated the polar magnetic flux for two sunspot pairs emerging on the two hemispheres,
to find out how much flux from the sunspots reaches the poles.
We calculate the polar flux by integrating $B_r$ over only those regions of the surface between
$60^ {\circ}$ latitude and the pole where $B_r$ has one sign (positive in the north pole). While positioning the sunspot pairs by hand using
the SpotMaker algorithm, we injected $1\times 10^{22}$ \rm{Mx} flux in each spot.
A normalized polar flux is estimated by dividing the signed flux  
by the input flux ($1\times10^{22}$ \rm{Mx}). In Figure~\ref{pflux_2}, we have shown the percentage of normalized polar flux with time
for the spot pairs emerging at different latitudes. It is evident from this figure that 
around $1.76 \%$ of the input flux can reach the pole when the spot pair emerges at a high latitude like $40^{\circ}$,
whereas $0.2 \%$ of the input flux can reach the pole when the spot pair is at a low latitude like $5^{\circ}$.
Keeping in mind that we have used an unrealistically high tilt of $40^{\circ}$, we point out that the flux reaching
the poles will be less for more realistic tilts. It is instructive to compare our result with relevant
observational data. \citet{Schrijver94} and \citet{Solanki02a} analyzed the 
NSO Kitt Peak magnetograph data and estimated the maximum active regions flux during solar maxima to be around $5 \times 10^{23}$ \rm{Mx}.
\citet{Munoz12} calibrated century long polar faculae data from Mount Wilson Observatory and estimated the time evolution of the polar flux, 
finding its maximum value to be around $1.5\times10^{22}$ \rm{Mx} for an average cycle. Although these values are not from 
a single dataset and many other observational constraints should be taken into account, a simple division of these values of
flux quoted above suggests that around $3\%$ of the sunspots flux can contribute in the
polar flux. Our theoretical model gives a value having the same order of magnitude, although our theoretical
values are a little bit on the lower side.

All the results presented so far for two sunspot pairs in different hemispheres were obtained by putting
both the pairs in the same longitude. This helped in magnetic fluxes of the two leading sunspots canceling
each other by diffusing across the equator. One important question is whether the final outcome will be
different if the two sunspot pairs in the two hemispheres are widely separated in longitude. Figure~\ref{sfield_3} shows
the surface evolution of magnetic flux in such a case, which can be compared with Figure~\ref{sfield_2}.
We find that the magnetic fluxes from the following sunspots in the two hemispheres are carried toward
the pole exactly as in Figure~\ref{sfield_2}.  However, the evolution of magnetic fluxes from the leading
sunspots is quite intriguing. Because of the gap in longitude, these fluxes cannot cancel with each other
across the equator so easily. However, these fluxes still diffuse across the equator, as seen in Figure~\ref{sfield_3}  ,
and, if we average over longitude, the averaged values are found to be virtually identical with the averaged
values that we get in the case of Figure~\ref{sfield_2}.  When we plotted figures similar to Figure~\ref{mfield_2} 
and Figure~\ref{rfield_2} for this case, they turned out to be indistinguishable from Figure~\ref{mfield_2} 
and Figure~\ref{rfield_2}.

\begin{figure}[!htbp]
\centerline{\includegraphics[width=0.4\textwidth,clip=]{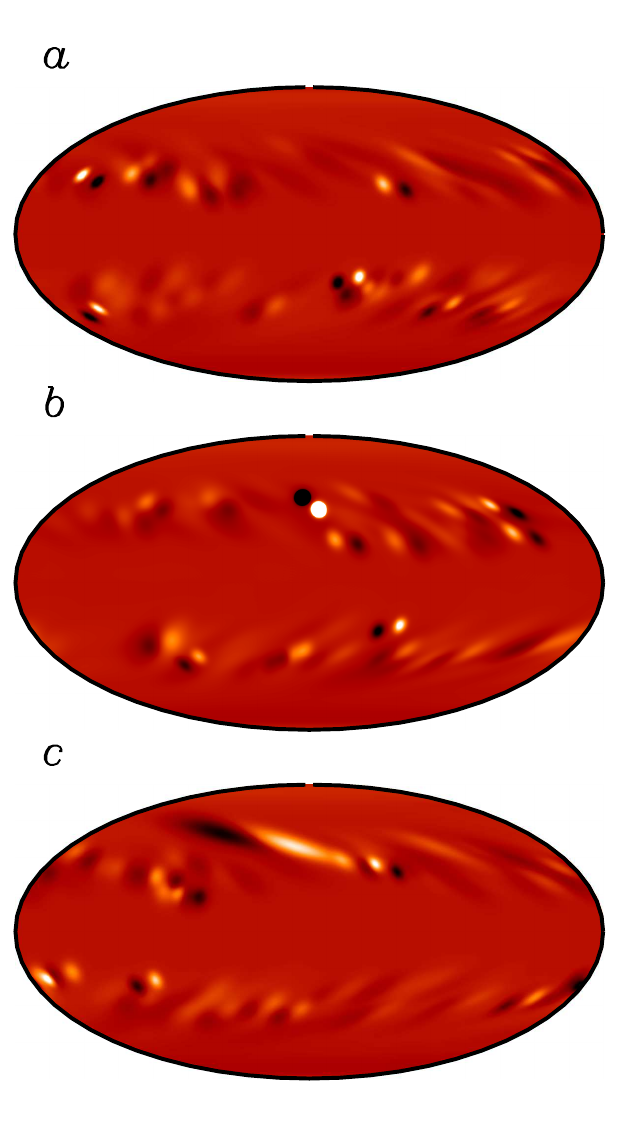}} 
\caption{Radial magnetic field structures are shown for the case when the ``anti-Hale" sunspot pair appears at $40^{\circ}$ latitude and at the middle phase of the cycle. (a) Prior to 3 months before the anti-Hale sunspot pair to be appeared, (b) During the emergence of anti-Hale sunspot pair and, (c) 3 months after the anti-Hale sunspot pair has emerged. Here white color shows the outward-going radial fields and black color represents inward-going radial fields. The color scale is set at $\pm100$ kG for all three cases.}
\label{ss_ahale}
\end{figure}

\begin{figure*}[!htbp]
\centerline{\includegraphics[width=1.0\textwidth,clip=]{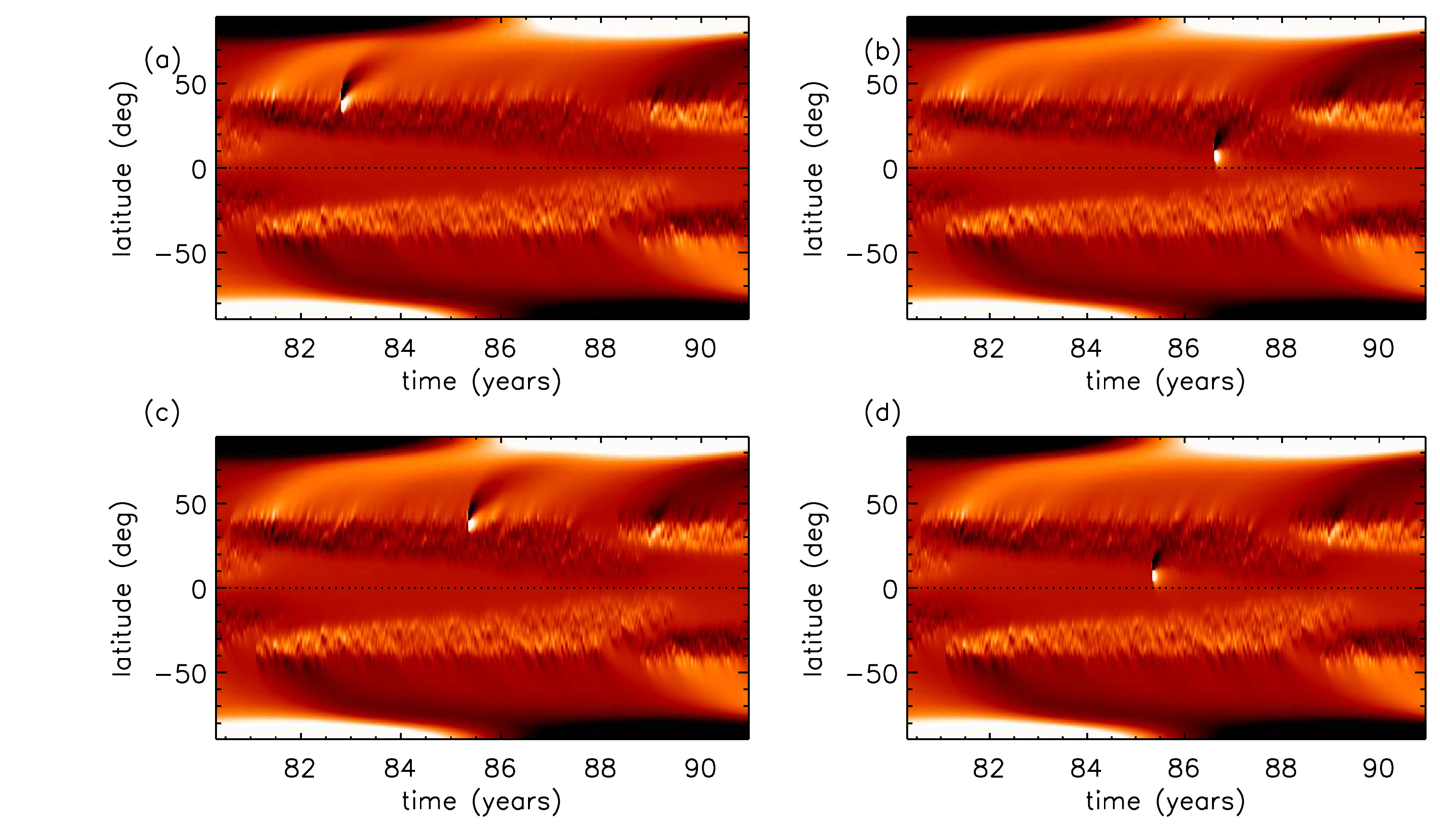}} 
\caption{Butterfly diagram with an ``anti-Hale" sunspot pair at different latitude and at different phase of the solar cycle. 
(a) At early phase of the cycle and at $40^{\circ}$ latitude. (b) Late phase of the cycle and at $10^{\circ}$ latitude. 
(c) Middle phase and at $40^{\circ}$ and (d) Middle phase and at $10^{\circ}$ latitude. Color scale is set at $\pm15$ kG for all four cases.}
\label{bfly_ahale}
\end{figure*}

\section{The contribution of bipolar sunspots not obeying Hale's law}

Joy's law for tilts of sunspot pairs is only a statistically average law.  We see a spread
of tilt angles around Joy's law. This spread is believed to be caused by the action of
turbulence on rising flux tubes \citep{Longcope02,weber11} and is one of the main sources
of irregularity in the solar cycle \citep{CCJ07, CK09,Chou13}.
It is well known that some bipolar sunspots appear with wrong magnetic polarities not 
obeying Hale's polarity law.  Because of the spread in tilt angles around Joy's law, 
it is certainly expected that a few outliers in this spread would violate Hale's law.
\citet{SK12} estimated that about 4\% of medium and large sunspots violate Hale's 
law---see their Figure 7.  Since the number of such sunspots is small, it is not 
surprising that due to statistical fluctuations, more of such sunspots 
violating Hale's law may appear in some particular cycles compared to other cycles. 
This fact assumes significance in the light of the suggestion made by \citet{Jiang15} on the basis of 
their SFT calculations that a few large ``anti-Hale'' sunspot pairs may significantly decrease
the strength of the polar field produced at the end of the cycle.  Especially, \citet{Jiang15}
suggested that the weak polar field at the end of cycle 23 was caused by a few prominent 
anti-Hale sunspot pairs present in that cycle.  In contrast, they argue that not too many such
anti-Hale sunspot pairs appeared in cycles 21 and 22, as a result of which such a decrease
of the polar field did not happen in those cycles.

Since we have seen that some insights gained from SFT calculations have to be modified---especially 
results connected with the build-up of the polar field---on the basis of more
realistic and complete 3D kinematic dynamo calculations, we now address the question whether
anti-Hale sunspot pairs have a large effect on the polar field even in 3D kinematic dynamo 
models. We now use our reference model presented in \S~3 and place a large anti-Hale sunspot
pair by hand to study its effect on the build-up of the polar field. To make its effect
visible, we take this anti-Hale sunspot pair to carry 25 times
the magnetic flux carried by the other regular sunspots and to have tilt angle $30^{\circ}$.
We can say that the tilt angle is $-30^{\circ}$, if we define the tilt angle by
following the convention that its value is positive for sunspot pairs obeying Hale's
law.
 
\begin{figure}[!htbp]
\centerline{\includegraphics[width=0.5\textwidth,clip=]{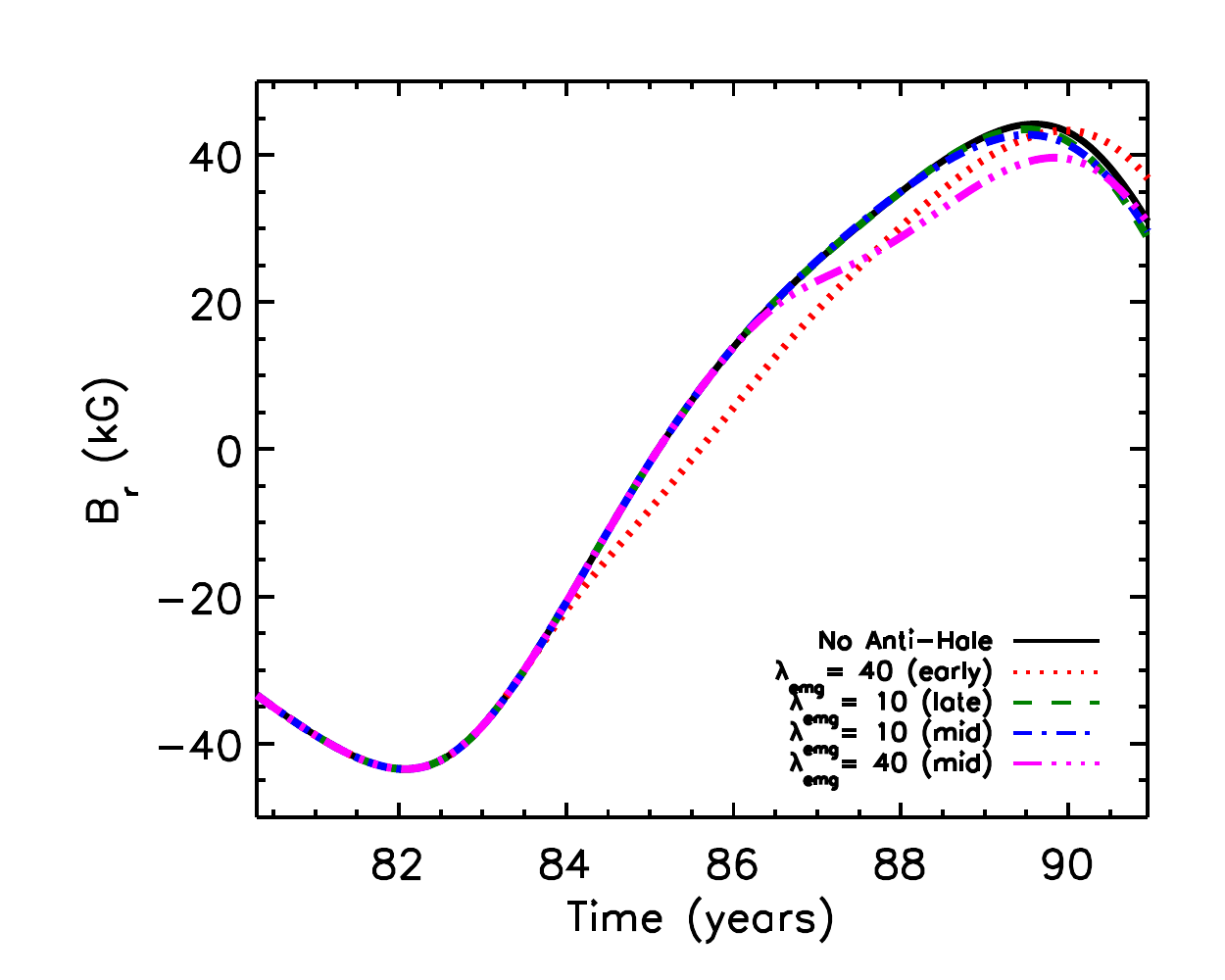}} 
\caption{Polar field evolution with time for one complete solar cycle with the ``anti-Hale" sunspot pair at different 
locations and different times of the cycle. Solid black line represents the regular cycle with no anti-Hale sunspot pair. 
Red dotted line indicates the poloidal field evolution with an anti-Hale pair at $40^{\circ}$ latitude at an early phase 
of the cycle. Green dashed line represents poloidal field with an anti-Hale pair at $10^{\circ}$ and late phase. Blue 
dashed and magenta long dashed lines indicate the poloidal field with an anti-Hale pair at middle of the cycle but 
at $10^{\circ}$ and at $40^{\circ}$ latitude, respectively.}
\label{pfield_ahle}
\end{figure}

We want to understand how the effect of the anti-Hale sunspot pair depends on the emergence
latitude, as well as the phase of the cycle, when it makes its appearance. So, we consider
four different cases. Since sunspots appear at high latitudes in the early phase of the
cycle and at low latitudes in the late phase, we consider one case by putting the anti-Hale
sunspot pair at the high latitude of $40^{\circ}$ in the early phase and another case
by putting the pair at the low latitude of $10^{\circ}$ in the late phase. The two other
cases considered involve putting the large anti-Hale sunspot pair at $40^{\circ}$ and
$10^{\circ}$ (in separate case studies) in the middle phase of the cycle. 
The radial fields on the surface in Mollweide projection is shown for
a case where an ``anti-Hale" sunspot pair is placed at $40^{\circ}$ latitude during the middle phase of the
cycle, in Figure~\ref{ss_ahale}.
Figure~\ref{bfly_ahale} shows how $B_r$ evolves in a time-latitude plot (a ``butterfly diagram'') for these four cases.
The effect on the polar field can be seen more clearly in Figure~\ref{pfield_ahle} where we plot the time evolution
of the polar field for these four cases, along with the reference case without an anti-Hale
sunspot pair.

It is clear from Figure~\ref{pfield_ahle} that even a very large anti-Hale sunspot pair placed at a low latitude
like $10^{\circ}$ does not have much effect on the polar field. Presumably, the opposite
fluxes from the two sunspots neutralize each other before they reach the poles. This becomes
quite apparent by looking at Figures~\ref{bfly_ahale}(b) and \ref{bfly_ahale}(d). We see that the sunspot pairs 
at low latitudes produce a kind of ``surge'' behind them, but it does not reach the poles. The effect
of anti-Hale pairs at higher latitudes is certainly much more pronounced. We see in Figures~\ref{bfly_ahale}(a)
and \ref{bfly_ahale}(c) that the surges behind these anti-Hale pairs reach the pole in these situations. If 
an anti-Hale sunspot pair appears at $40^{\circ}$ in the early phase of the cycle, then we
see in Figure~\ref{pfield_ahle} that the build-up of the polar field is weakened and delayed, but eventually
the polar field reaches almost the strength we would expect in the absence of the anti-Hale
sunspot pair.  However, when the anti-Hale sunspot pair is put at $40^{\circ}$ in the middle 
phase of the cycle, it is clear from Figure~\ref{pfield_ahle} that polar field can be reduced by about
17\%. But remember that we get this large reduction by assuming the anti-Hale sunspot pair to be unrealistically
large. Our conclusion is that anti-Hale sunspot pairs do
affect the build-up of the polar field---especially if they appear at high latitudes in
the middle phase of the cycle---but the effect does not appear to be very dramatic.
As for the suggestion of \citet{Jiang15} that the weakness of the polar field at
the end of cycle~23 was due to the appearance of several anti-Hale sunspot pairs, we feel that this is an interesting suggestion
which merits further detailed study in order to arrive at a firm conclusion.


\section{Conclusion}

Historically the evolution of the Sun's magnetic field with the solar cycle has been studied
extensively through two classes of 2D theoretical models: the 2D kinematic dynamo model and
the surface flux transport (SFT) model. We argue that the 3D kinematic dynamo model incorporates
the attractive aspects of both, while being free from the limitations of both.  On the one hand,
this model can treat the Babcock--Leighton mechanism more realistically in 3D, which is not
possible in the 2D kinematic dynamo model.  On the other hand, it includes the vectorial nature
of the magnetic field and various subsurface processes which are left out in SFT models.
\citet{Cameron12} have pointed out that the results of SFT model agree with the results of 2D flux
transport dynamo model on the inclusion of a downward pumping.  

In order to study the build-up of the Sun's polar field with a 3D kinematic dynamo model, we
first construct an appropriate self-excited model. The poloidal field generated by the Babcock--Leighton
mechanism near the solar surface has to be transported to the tachocline in order for the solar
dynamo to work.  This transport can be achieved in two ways: (i) due to advection by the
meridional circulation; or (ii) due to diffusion across the convection zone. There are reasons
to believe that (ii) is the appropriate transport mechanism inside the Sun. The earlier papers
by \citet{MD14} and \citet{MT16} presented self-excited dynamo models dominated by advection by
the meridional circulation. We believe that we are the first to construct self-excited 3D kinematic
dynamo model dominated by diffusion. We have briefly looked at the question of parity, although
the limitation of computer time prevented us from an exhaustive study of the subject.

We use this dynamo model to study how the polar field builds up from the decay of one tilted
bipolar sunspot pair and two symmetrically situated bipolar sunspot pairs in the two hemispheres.
We find that the polar field which arises from such sunspot pairs ultimately disappears due to
the emergence of poloidal flux at low latitudes and its subsequent subduction by the meridional flow.
This process is not included in the SFT models, in which the polar field can only be neutralized by
diffusion with a field of opposite polarity. So we conclude that SFT models do not capture the dynamics
of polar fields realistically and one has to be cautious in interpreting the SFT results pertaining
to polar magnetic fields. Our results differ most dramatically from the SFT results when we put two
symmetric bipolar sunspot pairs in the two hemispheres very near the equator. Then, the magnetic
fields of the two leading sunspots on the two sides of the equator cancel each other. At the same
time, the magnetic fields of the following sunspots which formed at higher latitudes are advected
by the meridional circulation to the poles, ultimately causing a magnetic dipole of the Sun. In
the SFT model, this is the whole story and we get an asymptotically steady dipole. In our 3D kinematic
model, on the other hand, magnetic field lines between the two hemispheres can get detached when
they are pulled by the meridional circulation in the opposite directions. As a result, radial magnetic
fields with signs opposite to the polar fields develop in the lower latitudes. This is not possible
in the SFT model in the absence of any source of radial magnetic field in the lower latitudes.
Finally, the detached magnetic loops in the two hemispheres are subducted underneath the surface
by the meridional circulation, contradicting the SFT result that the magnetic dipole of
the Sun would be asymptotically steady in this state. While the SFT models played a tremendously
important historical role in our understanding of how the magnetic field on the solar surface
evolves, we should keep in mind that these models cannot capture certain aspects of the dynamics of
the Sun's polar magnetic fields due the intrinsic limitations of these models.

Finally, we look into the provocative question of whether a few large sunspot pairs violating
Hale's law could have a large effect on the strength of the polar field. We find that such
anti-Hale sunspot pairs do produce some effect on the Sun's polar field---especially if they
appear at higher latitudes during the mid-phase of the solar cycle---but the effect is not 
very dramatic. The question of whether a few large anti-Hale sunspot pairs could be the principal
cause behind the weakness of the polar field at the end of some cycles, for example cycle~23, needs to
be analyzed carefully.

\medskip

The computations were performed on Yellowstone cluster provided by National Center for Atmospheric Research (NCAR)
and SahasraT cluster at Indian Institute of Science. We thank Bidya Binay Karak and an anonymous referee
for constructive comments which helped in improving the manuscript.
Partial support was provided from the JC Bose Fellowship awarded to A.R.C. by the Department
of Science and Technology, Government of India. G.H. thanks CSIR, India for financial support.

\bibliography{myref}
\end{document}